\documentclass[%
reprint,
superscriptaddress,
amsmath,amssymb,
aps,
prx,
]{revtex4-2}

\usepackage{chemformula} 
\usepackage{filecontents}
\usepackage[T1]{fontenc} 
\usepackage{amsmath} 

\usepackage{xr} 
\newcommand{\unit}[1]{\,\mathrm{#1}}
\newcommand{\degC}{^\circ{\mathrm C}}

\begin{document}

	\title{Intrinsic cell-to-cell variance from experimental single-cell motility data}
	
	
	\author{Anton Klimek}
	\affiliation{Fachbereich Physik, Freie Universit{\"a}t Berlin, 14195 Berlin, Germany}

	\author{Johannes C. J. Heyn}
	\affiliation{Physics Faculty and Center for NanoScience, Ludwig Maximilians Universit{\"a}t, 80539 M{\"u}nchen, Germany}
	
	\author{Debasmita Mondal}
	\affiliation{Department of Physics, Indian Institute of Science, 560012 Bangalore, India}
	\affiliation{James Franck Intitute, University of Chicago, 60637 Chicago, USA}
	
	\author{Sophia Schwartz}
	\affiliation{Fachbereich Chemie und Biochemie, Freie Universit{\"a}t Berlin, 14195 Berlin, Germany}
	
	\author{Joachim O. Rädler}
	\affiliation{Physics Faculty and Center for NanoScience, Ludwig Maximilians Universit{\"a}t, 80539 M{\"u}nchen, Germany}
	
	\author{Prerna Sharma}
	\affiliation{Department of Physics, Indian Institute of Science, 560012 Bangalore, India} 
	\affiliation{Department of Bioengineering, Indian Institute of Science, 560012 Bangalore, India}
	
	\author{Stephan Block}
	\affiliation{Fachbereich Chemie und Biochemie, Freie Universit{\"a}t Berlin, 14195 Berlin, Germany}
	
	\author{Roland R. Netz}
	\affiliation{Fachbereich Physik, Freie Universit{\"a}t Berlin, 14195 Berlin, Germany}
	\email{rnetz@physik.fu-berlin.de}

	\begin{abstract}
	
When analyzing the individual positional dynamics of an ensemble of moving objects, the extracted parameters that characterize the motion of individual objects,
such as the mean-squared instantaneous velocity or the diffusivity, 
exhibit a spread that is due to the convolution of three different effects:
i) Motion stochasticity, caused by the fluctuating environment and enhanced by limited observation time,
ii) measurement errors that depend on details of the detection technique, and 
iii) the intrinsic parameter variance that characterizes differences between individual objects,
the quantity of ultimate interest. 
We develop the theoretical framework to separate these effects using the generalized Langevin equation (GLE), which constitutes the most general description of active and passive dynamics, 
as it derives from the general underlying many-body Hamiltonian for the studied system without approximations.
We apply our methodology to determine intrinsic cell-to-cell differences of living human breast-cancer cells, algae cells and, as a benchmark, size differences of passively moving polystyrene beads in water. 
We find algae and human breast-cancer cells 
to exhibit significant individual differences, reflected by the spreading of the intrinsic mean-squared instantaneous velocity over two orders of magnitude, 
which is remarkable in light of the genetic homogeneity of the investigated breast-cancer cells and highlights their phenotypical diversity. 
Quantification of the intrinsic variance of single-cell properties is relevant for infection biology, ecology and medicine and opens up new possibilities to estimate population heterogeneity on the single-organism level in a non-destructive manner.
Our framework is not limited to motility properties but can be readily applied to general experimental time-series data.

	\end{abstract}

	\maketitle

	\section{Introduction}
	Cells of a population typically exhibit largely different genotypes and phenotypes, which creates optimized fitness in reaction to external stimuli. Oftentimes, it is important to know how heterogeneous cells or organisms are, e.g. in order to estimate survival probabilities in reaction to environmental changes \cite{levien2021non,conner1999effects,fox2006consequences} or in order to determine the likelihood of an infection by pathogens \cite{avraham2015pathogen,fleming2015effects}.
	The heterogeneity of a population of organisms is typically quantified by hand-picked parameters, for instance the cell size \cite{asadullah2021combined}, specific binding coefficients \cite{aird2003endothelial,rigler1999specific}, positional speed of the organism \cite{jerison2020heterogeneous,studenski2011gait,maiuri2015actin} etc.
	One experimentally relatively easily obtainable observable is the position of an organism measured over time, i.e. its positional trajectory, which requires minimal perturbation of the organism during the measurement and can be used to distinguish different organisms from each other \cite{klimek_data-driven_2024}.
	To quantify and compare trajectories one usually assumes a model.
	The most general equation of motion for a general observable that can be derived exactly from the underlying 
	Hamiltonian dynamics is the generalized Langevin equation (GLE), which was successfully used to model binary reaction dynamics in water \cite{Florian2022a}, 
	water vibrational IR line shapes \cite{Florian2022b}, butane dihedral dynamics \cite{dalton2024role},
	single cell motion \cite{mitterwallner_non-markovian_2020,klimek_data-driven_2024} and protein folding \cite{ayaz2021non,dalton2023fast}.
	The GLE includes non-Markovian effects, i.e., the memory of the trajectory of its past, and is valid even for non-equilibrium processes, such as cell motion \cite{klimek_data-driven_2024,roland_neq_2023}.
	Important limiting cases of the GLE are the persistent random walk (PRW) and Brownian diffusion models as well as active walk models such as the active Ornstein-Uhlenbeck model \cite{martin2021statistical_OU,mitterwallner2020negative}, which are explained in more detail further below.
	The motion of any organism and of any microscopic object, be it active or passive, contains intrinsic stochasticity due to the fluctuating environment \cite{viswanathan2011physics}, which is fully captured by the description with the GLE.
	
	Every trajectory that is observed over a finite time span inevitably leads to some uncertainty in the extracted parameters.
	Therefore, every experimental determination of model parameters contains a spread that originates from the stochasticity of the motion, entangled 
	 with experimental measurement errors and with 
	the actual spread due to differences between the individual moving objects.
	There are many methods available to estimate parameters of stochastic processes \cite{mitterwallner_non-markovian_2020,moon1996expectation} and to estimate the inevitable variance of such parameters \cite{law2007simulation,barton2001resampling}.
	One standard way to asses parameter uncertainties for stochastic processes is to simulate the stochastic model and to compare with experimental data \cite{law2007simulation,massada2008incorporating}.
	However, the field of motility analysis yet lacks methods to disentangle the spread of intrinsic properties between individual objects from the noise of the environment and other experimental uncertainties.
	
	In this paper we introduce an approach to tackle this problem: By simulating the stochastic process of motion for the experimental finite observation time with the universally applicable GLE model of motion and using a statistical variance analysis, we estimate the intrinsic spread of the parameters characterizing the individual moving objects.
	We show that the active and passive movements of individual objects of three very different types, chosen to cover a wide range of different patterns of motion, are all perfectly described by the GLE.
	Specifically, we consider the passive diffusion of polystyrene beads in water, the active motion of human breast-cancer cells on a substrate and the flagella-propelled motion of algae cells in two-dimensional confinement.
	The polystyrene-bead system constitutes a benchmark with known bead-to-bead differences, for which we confirm that our model estimates the correct intrinsic parameter variance.
	The algae and the human breast-cancer cell data reveal a similar large variance over two orders of magnitude in the intrinsic motion parameters, which is noteworthy as the cancer cells derive from a single cell line and thus share very similar genetic code, while the algae cells stem from multiple colonies and presumably are genetically rather heterogeneous.
	This showcases a significant phenotypic cell variance, manifested in the differences of the motion of individual cells, that is largely independent of the genetic diversity.
	Our framework is generally applicable to ensembles of time-series data and allows to separate stochastic environmental influences from intrinsic variations
	between individual ensemble members.

	\begin{figure*}
		\centering
		\includegraphics[width=\textwidth]{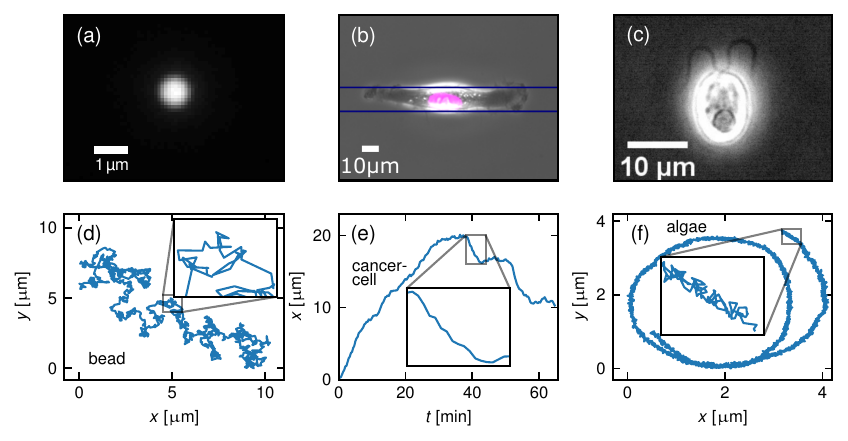}
		
		\caption{Microscopy image of (a) a fluorescently labeled polystyrene bead with radius $r=0.5\unit{\mu m}$ in three-dimensional solution, (b) a living breast-cancer cell (MDA-MB-231), where the pink area is the fluorescently labeled nucleus of the cell and the horizontal lines indicate the fibronectin-covered lane on which the cell moves, and (c) a Chlamydomonas reinhardtii algae cell confined between two glass plates at a separation of about $10\unit{\mu m}$. In (d)-(f) we show exemplary trajectories of the different moving objects shown in (a)-(c). The trajectory lengths are (d) $L=23.7\unit{s}$, (e) $L=64.7\unit{min}$ and (f) $L=8.2\unit{s}$, where the insets show enlarged sections of the trajectories.}
		\label{fig_0}
	\end{figure*}
	
	\section{Results}
	\subsection{Trajectory model}
	We analyze trajectories of three different moving objects, firstly the purely passive motion of polystyrene beads in water with a known size distribution, secondly the motion of human breast-cancer cells of line MDA-MB-231 on a one-dimensional micro lane \cite{schreiber2021adhesion,amiri2023multistability,mitterwallner_non-markovian_2020} and finally the motion of micro algae cells Chlamydomonas reinhardtii confined between two glass plates \cite{mondal_strong_2021,klimek_data-driven_2024}, which resembles their natural environment in soil. Exemplary microscopy images of the three different moving objects considered in the experiments are shown in Figs. \ref{fig_0}a-c.
	We choose these three systems as they cover important limiting scenarios, namely, passive Brownian diffusion (polystyrene beads), active persistent random motion (cancer cells) and strongly non-Markovian active motion (algae cells).
	Thus, our data sets include Markovian motion and non-Markovian motion in equilibrium as well as out of equilibrium.
	The experiments yield two-dimensional trajectories $x(t), y(t)$ for the center position of the polystyrene beads and the algae cells, examples of which are shown in Figs. \ref{fig_0}d,f, respectively, and one-dimensional trajectories $x(t)$ for the breast-cancer cells, an example of which is shown in Fig. \ref{fig_0}e.
	We describe all trajectories by the GLE
	\begin{equation}
		\label{eq_gle}
		\ddot{x}(t) = -\int_{t_0}^{t} \Gamma_v(t-t') \dot{x}(t') dt' + F_R(t) 
	\end{equation}
	with an identical equation for $y(t)$ in the case of two-dimensional motion.
	The range of different types of motion in the experiments illustrates the universal applicability of the GLE.
	For Gaussian motion processes, which perfectly describe the experiments as explained further below, the equation of motion is linear and there is no coupling between the motion
	in $x$ and $y$ direction, therefore, we average all two-dimensional trajectory data over the two directions.
	In the GLE eq. \eqref{eq_gle}, $\ddot{x}(t) = \dot v(t)$ denotes the acceleration of the position, $\Gamma_v(t)$ is a memory kernel that 
	describes how the acceleration at time $t$ depends on the velocity $\dot{x}(t')= v(t')$ at previous times 
	and therefore accounts for non-Markovian friction effects, and 
	$F_R(t)$ is a random force that describes interactions 
	with the surrounding and the interior of the moving object.
	We show further below that all different types of active and passive motion we consider are perfectly described by the GLE eq. \eqref{eq_gle}.
	Since the experimental systems are isotropic and homogeneous in the observed space, no 
	deterministic force term appears in the GLE.
	In fact, the GLE in eq. \eqref{eq_gle} is the most general equation of motion for Gaussian unconfined motion processes and can be derived by projection from the underlying many-body Hamiltonian, even in the presence of non-equilibrium effects, which obviously are present for living organisms 
	\cite{mori_transport_1965,zwanzig1961memory,cihan2022_hybrid_gle,roland_neq_2023}.
	An important special case of the GLE eq. \eqref{eq_gle} is the Markovian limit with $\Gamma(t) = 2 \delta(t) / \tau_m$, which leads to the well known Langevin equation and describes persistent motion. In the so-called overdamped limit, in which case the persistence time $\tau_m$ goes to zero, one recovers the even simpler Brownian diffusion model.
	
\begin{figure*}
		\centering
		\includegraphics[width=\textwidth]{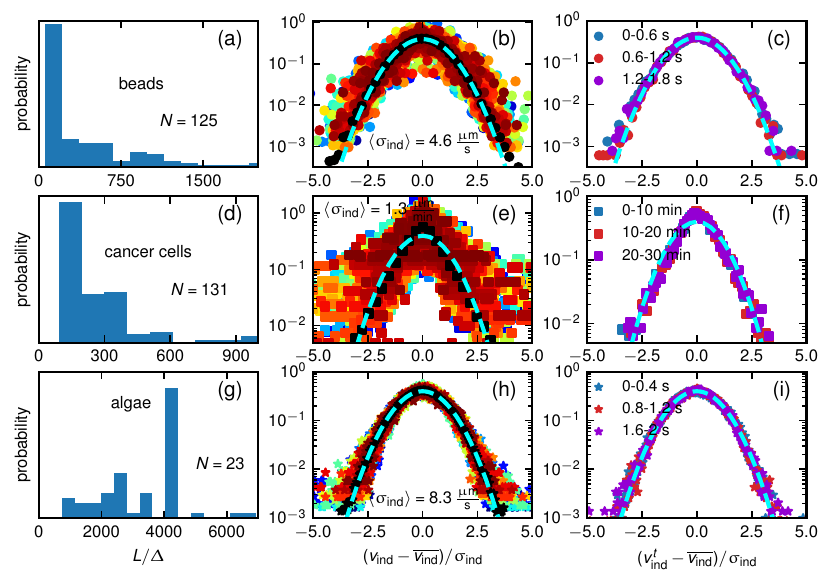}
		
		\caption{Distributions of trajectory lengths $L$ in units of the experimental discretization time $\Delta$ for (a) polystyrene beads with $\Delta=0.02\unit{s}$, (d) breast-cancer cells with $\Delta=20\unit{s}$ and (g) algae cells with $\Delta=0.002\unit{s}$. Velocity distributions of individual moving objects rescaled by subtracting their individual mean velocity $\overline{v_{\rm{ind}}}$ and dividing by their individual standard deviation $\sigma_{\rm{ind}}$ for (b) polystyrene beads, (e) breast-cancer cells and (h) algae cells, where individual cells are distinguished by color and population mean values $\langle \sigma_{\rm{ind}}\rangle$ are given in the plots. The black symbols in (b),(e),(h) represent the population average over all moving objects, which agrees very well with the standard normal distribution shown as a dashed blue line. The average over individually rescaled velocity distributions is identical for three different time windows for (c) polystyrene beads (f) breast-cancer cells and (i) algae cells, which suggests stationarity.}
		\label{fig_1}
	\end{figure*}

	If the motion can be described as a Gaussian process,
	which for polystyrene beads, cancer cells and for algae cells is suggested by the fact that the single-individual velocity distributions 
	are Gaussian, as shown in Figs. \ref{fig_1}b,e,h, the random force is Gaussian as well \cite{roland_neq_2023} with correlations given by 
	\begin{equation}
		\label{eq_fdt_neq}
		\langle F_R(t) F_R(0) \rangle = \Gamma_R (t) \,,
	\end{equation}
	where $\Gamma_R(t)$ denotes the symmetric random-force kernel.
	Averaging cell velocities over the cell population can result in non-Gaussian velocity distributions, even though distributions are Gaussian on the individual trajectory level \cite{klimek_data-driven_2024,mitterwallner_non-markovian_2020}, as we show in the SI.
	Therefore we check the Gaussianity on the single-trajectory level in Fig. \ref{fig_1}.
	
	The GLE eq. \eqref{eq_gle} does not contain time-dependent parameters and, therefore, describes a stationary process. Indeed, for the polystyrene beads, the cancer cells as well as for the algae cells the velocity distributions do not change in time as shown in Figs. \ref{fig_1}c,f,i, which suggests that the GLE can be used to model the objects' motion.
	
	For an equilibrium system, such as the passively moving polystyrene beads in water, the fluctuation dissipation theorem (FDT)
	predicts $\Gamma_R(t)=B\Gamma_v(\vert t\vert )$ with the mean-squared velocity $B\equiv\langle v^2\rangle$ given by
	$B=k_BT/m$ according to the equipartition theorem,
	where $m$ is the mass of the moving object and $k_BT$ denotes the thermal energy
	\cite{mori_transport_1965,zwanzig1961memory}. 
	For living cells, both FDT and equipartition theorem do not hold in general
	and thus there is no a priori reason why $B\Gamma_v(\vert t\vert )$ and $\Gamma_R(t)$ should be equal
	\cite{mizuno2007nonequilibrium,netz2018non_equilibrium}.
	Nevertheless, for non-equilibrium scenarios with Gaussian statistics one can construct a surrogate model 
	with an effective kernel $\Gamma(\vert t\vert)=\Gamma_R(t)/B=\Gamma_v(\vert t \vert)$ 
	that exactly reproduces the dynamics described
	by the non-equilibrium GLE with $ \Gamma_R(t) \neq B\Gamma_v(\vert t\vert)$ \cite{klimek_data-driven_2024,roland_neq_2023}.
	Thus, for each combination of $\Gamma_R(t)$ and $\Gamma_v(t)$, a unique $\Gamma(t)$ exists, which can be extracted from trajectories via the velocity autocorrelation function (VACF) \cite{mitterwallner_non-markovian_2020,klimek_data-driven_2024} (as shown in the SI).
	Since the VACF completely determines the dynamics of a Gaussian system \cite{roland_neq_2023b,klimek_data-driven_2024},
	this implies that the extracted 
	effective kernel $\Gamma(t)$ not only describes the VACF exactly but also completely characterizes the motion
	\cite{roland_neq_2023b}.
	Recent work, where the non-equilibrium GLE is derived from a suitably chosen time-dependent Hamiltonian,
	shows that for Gaussian non-equilibrium observables the condition $\Gamma_R(t) = B\Gamma_v(\vert t\vert)$ 
	is actually satisfied \cite{roland_neq_2023}, in line with our method that is based on extracting an effective kernel
	$\Gamma(t)$ for active cell motion.
	Since we show that the GLE is appropriate to model our experimental trajectory data, we can use it to produce synthetic trajectories with the same length as experimental trajectories shown in Figs. \ref{fig_1}a,c,g and compare them to the experiments, which is key to determining the variance of parameters characterizing the individual objects' differences.

	\subsection{Trajectory analysis}
\begin{figure*}
		\centering
		\includegraphics[width=\textwidth]{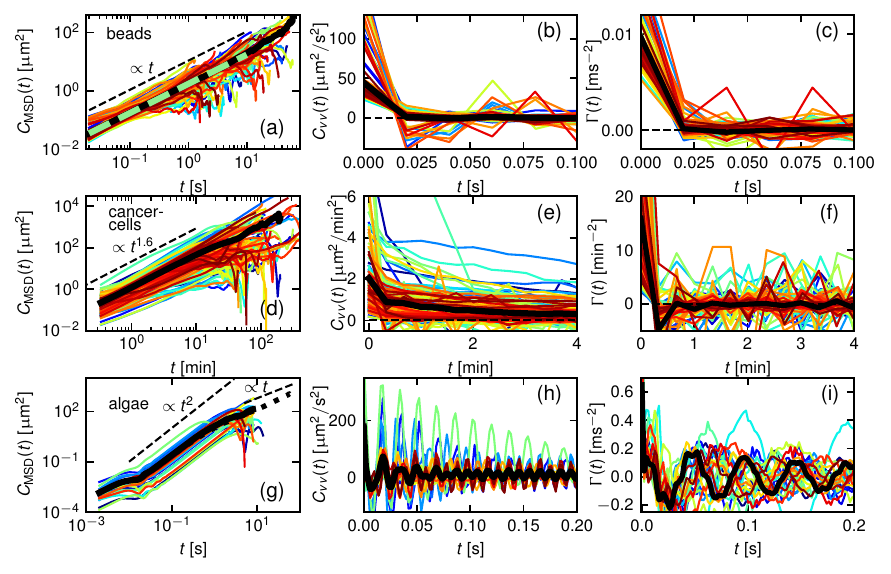}
		
		\caption{Results for the MSD, $C_{\textrm{MSD}}(t)$ defined in eq. \eqref{eq_def_msd}, 
			the VACF, $C_{vv}(t)$ defined in eq. \eqref{eq_def_cvv} and the friction kernel $\Gamma(t)$ extracted from the trajectories for polystyrene beads in (a)-(c), for cancer cells in (d)-(f) and for algae cells in (g)-(i).
			Different colors represent results for individual moving objects. 
			The black lines in the first two columns denote the average over all objects. 
			In the third column the black line denotes the friction kernel calculated from the average VACF. The dashed green line in (a) is the theoretical prediction for the diffusion of a sphere with radius $r=0.5\unit{\mu m}$ in water predicted by the Einstein relation eq. \eqref{eq_stokes_einstein}. The black dashed lines in the MSD plots indicate the scaling behavior. The dotted line in (g) shows the long time diffusive regime from a recording with lower spatial resolution and longer observation times.}
		\label{fig_2}
	\end{figure*}

	A suitable characterization of stochastic trajectories employs the VACF or the MSD, defined by
	\begin{align}
		C_{vv}(t) &= \langle v(0)v(t) \rangle \label{eq_def_cvv}\,,\\
		C_{\textrm{MSD}}(t) &= \langle (x(0)-x(t))^2 \rangle \label{eq_def_msd}\,,
	\end{align}
	respectively. Even though the VACF is just the curvature of the MSD, $C_{vv}(t) = \frac{1}{2}\frac{d^2}{dt^2}C_{\textrm{MSD}}(t)$, these correlation functions accentuate different aspects of trajectories.
	For the passively moving polystyrene beads, the MSD in Fig. \ref{fig_2}a demonstrates the purely diffusive nature of the motion, where the average over all beads (black symbols) agrees well with the Brownian prediction $C_{\rm{MSD}}(t)=2Dt$ for a sphere of radius $r=0.5\unit{\mu m}$ shown as a dashed green line.
	For this prediction we use the Einstein relation between the friction of a moving object with its environment $\gamma$ at temperature $T$ and the diffusivity $D$, where the friction of a sphere with radius $r$ depends on the solution viscosity $\eta$, according to
	\begin{equation}
		\label{eq_stokes_einstein}
		D = \frac{k_BT}{\gamma} = \frac{k_BT}{6\pi\eta r}\,.
	\end{equation}
	Velocities follow from trajectories using finite differences (see Methods eq. \eqref{eq_central_diff_vel}), which depend on the recording time step $\Delta$ and localization noise.
	The VACF for the polystyrene beads in Fig. \ref{fig_2}b exhibits a single peak at time zero, which reflects that consecutive displacements are completely uncorrelated and shows that the persistence time $\tau_m$ is much smaller than $\Delta$.
	This implies that the friction kernel $\Gamma(t)$ is a delta function, which perfectly matches the extracted kernel shown in Fig. \ref{fig_2}c.
	We explain the extraction scheme of the friction kernel in detail in the SI.
	For the actively moving cancer cells, the average MSD over all cells exhibits superdiffusive behavior $\propto t^{1.6}$ with a slightly decreasing slope for longer times shown in Fig. \ref{fig_2}d, where the long time diffusive regime is not well resolved due to larger noise (caused by less averaging) for long times.
	The VACFs of the actively moving cancer cells are decaying on the scale of a few minutes, as shown in Fig. \ref{fig_2}e.
	Similarly to the polystyrene beads, the friction kernels shown in Fig. \ref{fig_2}f exhibit a delta peak at time zero, followed by a dip at the first time step $t=\Delta=20\unit{s}$ and is then zero.
	This dip, which is also present for the algae cell kernels at the first time step $t=\Delta=0.002\unit{s}$ in Fig. \ref{fig_2}i, originates from localization noise (see SI for more information), as this noise reduces the correlation of consecutive velocities and influences the extracted kernel \cite{klimek_data-driven_2024}.
	For algae cells the MSD exhibits a long ballistic regime up to a few seconds followed by the long time diffusive regime, as shown in Fig. \ref{fig_2}g, whereas the VACF in Fig. \ref{fig_2}h shows strong oscillations that correspond to the flagella beat cycle with a frequency of $\sim 50\unit{Hz}$.
	The extracted friction kernel shown in Fig. \ref{fig_2}i shows oscillations as well, but the oscillations exhibit a slightly different frequency because of the complex dependence of the VACF frequency on the kernel amplitude and frequency \cite{klimek_data-driven_2024}.
	
	The extracted friction kernels imply different modes of motion.
	For the algae cells in Fig. \ref{fig_2}i, the kernel exhibits a delta peak at time zero followed by a decaying oscillation, which was previously modeled by \cite{klimek_data-driven_2024}
	\begin{equation}
		\label{eq_kern_osc}
		\Gamma(t)=2a\delta(t) + be^{-t/\tau}\left(\cos(\Omega t) + \frac{1}{\tau\Omega}\sin(\Omega t)\right) \,.
	\end{equation}

	The extracted kernels of the cancer cells in Fig. \ref{fig_2}f suggest a simple persistent random walk \cite{selmeczi2005cell} (note that an earlier study modeled a different cancer cell line by a slightly more complex model including an exponential decay in the friction kernel $\Gamma(t)$ \cite{mitterwallner_non-markovian_2020}).
	The persistent random walk is described by the GLE eq. \eqref{eq_gle} with
	\begin{equation}
		\label{eq_kern_prw}
		\Gamma(t) = 2 \delta(t) / \tau_m\,,
	\end{equation}
	where the transition from the ballistic to the long time diffusive regime, i.e. the persistence time $\tau_m$, occurs for cancer cells within the experimentally observed time scales; the mean value $\tau_m\approx 2\unit{min}$ is in the order of minutes and thus much larger than the time step of $\Delta=20\unit{s}$.
	Therefore, the observed superdiffusive scaling behavior of the ensemble-averaged MSD in Fig. \ref{fig_2}d is explained by the transition from ballistic $\propto t^2$ to the long time diffusive regime $\propto t$, which is confirmed by fits of the model to single-cell data explained below.
	In contrast to the cancer cells, the persistence time of the polystyrene beads estimated by $\tau_m=m/\gamma\approx 0.1\unit{\mu s}$ is much smaller than the experimental time resolution $\Delta=0.02\unit{s}$, so that only the long-time diffusive behavior is observed.
	Hence, we model the polystyrene beads by pure diffusion, in which case the GLE eq. \eqref{eq_gle} simplifies to
	\begin{equation}
		\label{eq_gle_diffusion_lim}
		\dot{x}(t) = \sqrt{D}\xi(t)\,,
	\end{equation}
	where $\xi(t)$ is uncorrelated white noise with $\langle \xi(t) \rangle=0$ and $\langle \xi(0) \xi(t) \rangle=2\delta(t)$.
	
\begin{figure*}
		\centering
		\includegraphics[width=\textwidth]{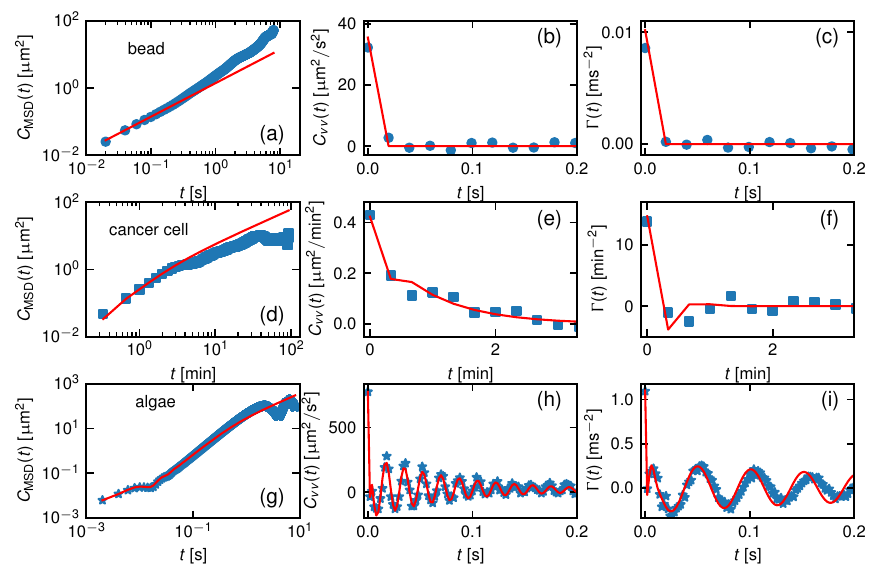}
		
		\caption{Experimental results for the MSD, $C_{\textrm{MSD}}(t)$, VACF, $C_{vv}(t)$, and friction kernel $\Gamma(t)$, of a single bead {(a)-(c)}, a single cancer cell {(d)-(f)} and a single algae cell {(g)-(i)} (blue symbols).
		The red lines denote predictions of the model defined in eq. \eqref{eq_gle_diffusion_lim} fitted to the bead MSD, of the model defined in eqs. \eqref{eq_gle} and \eqref{eq_kern_prw} fitted to the cancer cell VACF and of the model defined in eqs. \eqref{eq_gle} and \eqref{eq_kern_osc} fitted to the algae cell VACF.
		The individual bead, cancer cell and algae were picked at random; we show the individuals with the best fits to the models and the distribution of $R^2$, which measures the goodness of the fit, in the SI.}
		\label{fig_3}
	\end{figure*}

	In Fig. \ref{fig_3} we show fits of the analytical models to the MSD, VACF and friction kernel data using for a single polystyrene bead, eq. \eqref{eq_gle_diffusion_lim}, for a single breast-cancer cell eqs. \eqref{eq_gle} and \eqref{eq_kern_prw}, and for a single algae cell, eqs. \eqref{eq_gle} and \eqref{eq_kern_osc}, all randomly selected from the ensemble of moving objects.
	The GLE describes the data perfectly, which confirms the applicability of the GLE to describe the active or passive motion of single moving objects.
	Moreover, the model that includes localization noise (described in the Methods) correctly captures the decorrelation of consecutive velocities seen in the VACF in Figs. \ref{fig_3}e,h and the dip in the kernel at the first time step seen in Figs. \ref{fig_3}f,i. In the case of pure diffusion, consecutive displacements are uncorrelated, thus, the effect of the localization noise is not visible as strongly in the VACF and the friction kernel of the diffusing polystyrene beads in Figs. \ref{fig_3}b,c.

	\subsection{Reproducing experimental parameter distributions by simulation}
\begin{figure*}
		\centering
		\includegraphics[width=\textwidth]{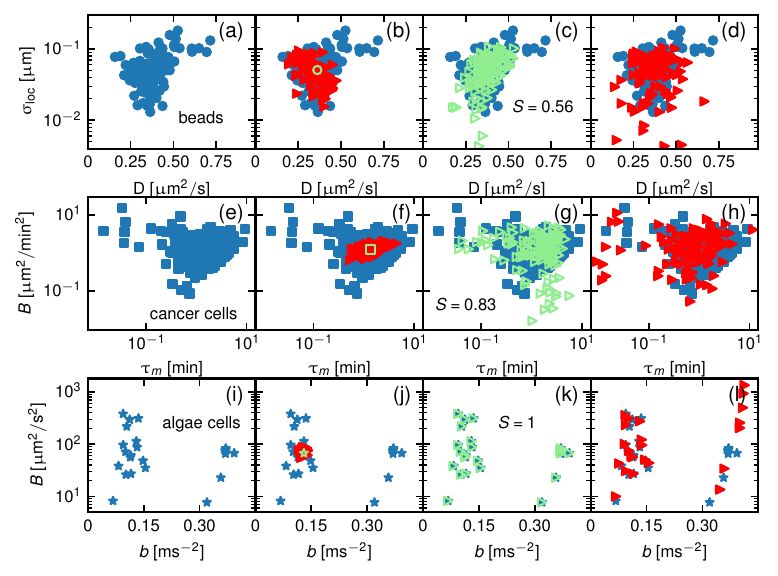}
		
		\caption{Distributions of GLE model parameters extracted from the experiments are shown as blue symbols, for polystyrene beads we show the localization noise width $\sigma_{\rm{loc}}$ and diffusivity $D$ in (a)-(d), for cancer cells the mean squared velocity $B$ and the persistence time $\tau_m$ in (e)-(h) and for the algae cells the mean squared velocity $B$ and the kernel oscillation amplitude $b$ (see eq. \eqref{eq_kern_osc}) in (i)-(l).
		The green symbols in (b), (f), (j) represent the respective median of the distribution of blue symbols and the red filled triangles represent the parameters extracted from a simulation where all objects have the median parameter set and using the experimental time step and length distribution, shown in Figs. \ref{fig_1}a,d,g.
		Empty triangles in (c), (g) denote Gaussian distributions of simulation parameters with $S=0.56$ and $S=0.83$, respectively, which correspond to our estimate for the actual parameter spread $S^*$ shown in Fig. \ref{fig_barplot}a.
		In (k) simulation parameters (green triangles) correspond to the parameters extracted from the experiment, which means that $S=1$.
		Simulations using the empty green triangle parameters in (c), (g), (k) and the respective model suggested by the data eqs. \eqref{eq_kern_osc}, \eqref{eq_kern_prw}, \eqref{eq_gle_diffusion_lim} and the experimental trajectory lengths lead to the filled red triangles in (d), (h), (l), which by definition of $S^*$ have the same covariance as the blue symbols.}
		\label{fig_4}
	\end{figure*}

	Even though all polystyrene beads are nearly identical, with the standard deviation of the radius being $\Delta r \sim 25\unit{nm}$ according to the manufacturer and as confirmed by our measurement of the bead radii by atomic force microscopy (explained in the SI), the extracted MSDs, VACFs and friction kernels of single beads scatter significantly around the mean, as seen in Figs. \ref{fig_2}a-c.
	Hence, the fitted single-bead parameters show a large spread, seen in Fig. \ref{fig_4}a, which does not reflect the true parameter spread that the beads actually exhibit, as we demonstrate further below.
	
	For living organisms it is often not known how similar individuals of one population are to each other. Naturally, the question arises: How much of the spread of the parameters extracted from experimental data originates from differences among individuals and how much comes from the motion stochasticity and experimental errors?
	
	We imagine the scenario that all individuals are completely identical and are represented by the median of the extracted parameter distribution.
	Here, we use the median instead of the mean, because a least-squares fit of a noisy exponential function leads to a log-normal distribution of the fitting parameters with the median being the best representation of the true parameter value, as explained in the SI.
	We simulate the same number of experimentally recorded trajectories $N$, with the same trajectory length distribution as in the experiment shown in Figs. \ref{fig_1}a,d,g using the median parameters estimated from the experimental trajectories using the GLE model at a temporal resolution $h$ which is 20 to 200 times smaller than the experimental discretization $\Delta$.
	We then use every $(\Delta/h)$-th point of the simulation to obtain a trajectory with time step $\Delta$ and add the localization noise to the trajectories, which is estimated directly by our fit described in the Methods.
	From such a simulation of $N$ trajectories with identical parameters, we extract the VACF and perform the same fit that we used to extract the parameters from the experiment.
	If the spread of the distribution extracted from the $N$ simulations with identical GLE parameters is comparable to the spread of the distribution extracted from the experiment, then it is likely that all moving objects are actually completely identical in their motion parameters.
	
	In fact, the experimentally determined spread of the single breast-cancer cell parameters shown in Fig. \ref{fig_4}e and of the single algae cell parameters shown in Fig. \ref{fig_4}i are much larger than the spread of the respective parameters extracted from simulations with identical input parameters, shown in Figs. \ref{fig_4}f,j by the red triangles.
	This suggests that the ensemble of cancer cells as well as the ensemble of algae cells are not characterized by identical parameters of motion.
	In contrast, in Fig. \ref{fig_4}b the spread of the bead parameter distribution extracted from the simulations with identical input parameters is almost as large as the spread of the parameter distribution extracted from the experiment, indicating that most of the observed spread originates from the finite length and the experimental noise, in line with the small variance of the bead radii determined by our AFM measurement.
	
	The simulation results in Figs. \ref{fig_4}f,j suggest that the individual cancer and algae cells are characterized by non-identical GLE parameters within the respective populations.
	To determine the actual parameter distribution of individual cells, we test whether simulations using input parameters from a Gaussian distribution with a certain covariance $\rm{Cov}^{\rm{inp}}$ result in an extracted parameter distribution that agrees in the covariance $\rm{Cov}^{\rm{sim}}$ with the covariance of the parameter distribution extracted from the experiment $\rm{Cov}^{\rm{exp}}$.
	For this we use a so-called permutation test that compares the covariances $\rm{Cov}^{\rm{sim}}$ and $\rm{Cov}^{\rm{exp}}$, which is explained in the Methods.
	
	We denote the summed ratio of the $d$-dimensional input parameter covariance for the simulation, $\rm{Cov}^{\rm{inp}}$, and the covariance of the distribution extracted from the experiment, $\rm{Cov}^{\rm{exp}}$, as
	\begin{equation}
		\label{eq_result}
		S = d^{-2} \sum_{i, j}^d \rm{Cov}^{\rm{inp}}_{ij} / \rm{Cov}^{\rm{exp}}_{ij} \,,
	\end{equation}
	where the covariance of a distribution of parameter vectors $\vec{z}$ is defined in the standard way as ${\rm{Cov}_{ij} = \langle (z_i-\langle z_i \rangle) (z_j-\langle z_j \rangle) \rangle}$ with $z_i$ being the components of the vector $\vec{z}$ and $\langle z_i \rangle$ the $d$ components of the mean vector, $d$ denotes the number of parameters.
	We consider input covariances in the range from $S=0$, which corresponds to all moving objects described by identical input parameters of motion, to $S=1$, where we use the parameter distribution extracted from the experiment as the simulation input parameters.
	For intermediate values $0<S<1$, simulation input distributions are drawn from Gaussian distributions with covariances that are uniformly rescaled by $S$.
	Exemplary simulation input parameter distributions are shown as green symbols in Figs. \ref{fig_4}c,g,k.
	
\begin{figure}
		\centering
		\includegraphics[width=0.53\textwidth]{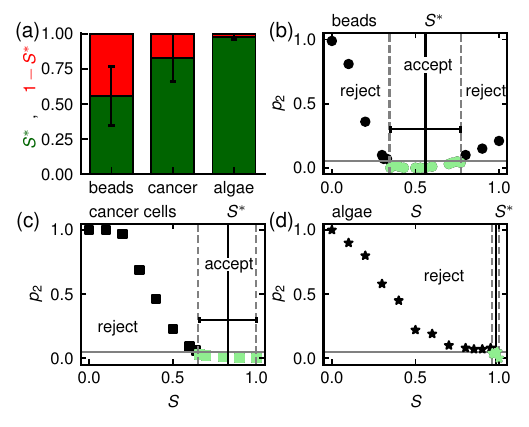}
		
		\caption{(a) Estimated covariance ratio $S^*$ of the intrinsic to the experimentally observed parameters shown in green. Likelihood measure $p_2$ that the covariance extracted from simulation $\rm{Cov}^{\rm{sim}}$ agrees with the experimental covariance $\rm{Cov}^{\rm{exp}}$ according to our statistical test described in the Methods as a function of $S$ (different $S$ originate from different simulation input covariances $\rm{Cov}^{\rm{inp}}$) for (b) beads, (c) cancer cells, (d) algae.
		The acceptance level of $p_2^*=0.05$ is indicated as a gray horizontal line and points below the line correspond to accepted $S$ values and are shown in green. The acceptance region is indicated by dashed vertical lines with the center, denoted by $S^*$, and the width defines the error bars shown in (a).}
		\label{fig_barplot}
	\end{figure}

	The main result of our method is the value $S^*$, for which the covariance of the distribution extracted from the experiment $\rm{Cov}^{\rm{exp}}$ agrees with the covariance of the distributions extracted from simulations $\rm{Cov}^{\rm{sim}}$ with the given input covariance $\rm{Cov}^{\rm{inp}*}$.
	Hence, we call $S^*$ the estimated covariance ratio and the corresponding input covariance $\rm{Cov}^{\rm{inp}*}$ the intrinsic parameter covariance of individual objects.
	Examples of distributions that are extracted from simulations with the green input distributions $\rm{Cov}^{\rm{inp}*}$ in \ref{fig_4}c,g,k are shown in Figs. \ref{fig_4}d,h,l (red triangles), which agree with the respective experimental distribution (blue symbols) according to our statistical test described in the Methods.
	The statistical test that we use compares the covariance of the experimentally determined parameter covariance $\rm{Cov}^{\rm{exp}}$ to the covariance of a parameter distribution extracted from a simulation $\rm{Cov}^{\rm{sim}}$ and yields a measure of how likely the two distributions originate from the same input covariance, which we call $p_2$.
	If several values of $S$ are accepted by our statistical test, meaning they lead to agreeing covariances $\rm{Cov}^{\rm{sim}}$ and $\rm{Cov}^{\rm{exp}}$ according to the test, we choose the midpoint of the accepted range as $S^*$ and use the upper and lower acceptance bounds to estimate the error of $S^*$, as demonstrated in Fig. \ref{fig_barplot}.

	As an example, the diffusion of polystyrene beads in water is described by the two-dimensional parameter vector comprising the diffusivity $D$ and the localization noise width $\sigma_{\rm{loc}}$, $\vec{z}=(D, \sigma_{\rm{loc}})$, the latter defines the uncertainty of the position data (see Methods).
	If the covariance of a parameter distribution extracted from the experiment was given by $\rm{Cov}^{\rm{exp}} = \begin{pmatrix}
		1 & 0.5\\
		0.5 & 1
	\end{pmatrix}$ with a median of $\vec{z}^{\rm{med}} = (0, 1)$, simulation input parameters for $S=0.1$ would be drawn from a Gaussian distribution with covariance $\rm{Cov}^{\rm{inp}} = \begin{pmatrix}
		0.1 & 0.05\\
		0.05 & 0.1
	\end{pmatrix}$ and median $ \vec{z}^{\rm{med}} =(0, 1)$ (here we omit units for clarity).
	As mentioned earlier, we use the median for generating simulation input parameters, because it is a better representation of typical parameters than the mean when using least square fits (see SI).
	If the parameter distribution covariance extracted from the simulation with $S=0.1$ $\rm{Cov}^{\rm{sim}}$ is the same as the parameter covariance extracted from the experiment $\rm{Cov}^{\rm{exp}}$ according to our statistical test, we do the same for $S=0.2$.
	Then we repeat the test increasing $S$ in steps of $0.1$ up to finally using $S=1$.
	In the regime of accepted $S$ values we use smaller steps to determine $S^*$. For accepted values in the range $S=0.05-0.15$, we would obtain the estimated covariance ratio as $S^*=0.1\pm0.05$.
	The evaluation of $S^*$ is graphically depicted in Figs. \ref{fig_barplot}b-d.

	For the polystyrene beads we find $S^* = 0.56\pm0.21$ as shown in Fig. \ref{fig_barplot}a, which translates into an estimate for the standard deviation of the bead radius $\Delta r=86\pm63\unit{nm}$ via the Einstein law eq. \eqref{eq_stokes_einstein} and using the viscosity of water at room temperature $\eta\approx 1\unit{mPs}$.
	The bead radius standard deviation of $\Delta r \sim 25\unit{nm}$ listed by the manufacturer and confirmed by our atomic force microscopy (AFM) measurements is close to the lower bound of our result.
	Nevertheless, it should be noted that other experimental factors such as the bead-surface properties contribute to additional individual-to-individual variance in diffusivity, thereby increasing the variance of the hydrodynamic radius $\Delta r$, which explain our relatively large estimate of $\Delta r$.
	
	For the cancer cells the estimated covariance ratio is given by $S^*=0.83\pm0.17$ shown in Fig. \ref{fig_barplot}a, which indicates a spread over nearly two orders of magnitude in the mean squared cell speed $B$ and in the persistence time $\tau_m$, seen in Fig. \ref{fig_4}g, even though all cells are from the same cell line and supposedly share the same genetic code.
	This demonstrates the large phenotype variance among cancer cells of the same cell line.
	The distribution extracted from the algae trajectories represents the true cell-to-cell variance accurately, as seen by comparing Figs. \ref{fig_4}k,l, and as reflected by $S^*=0.98\pm0.02$ shown in Fig. \ref{fig_barplot}a.
	
	Additionally to decomposing the spread on the individual-to-individual level, our approach allows us to estimate how long experimental recordings have to be, in order to obtain a certain level of $S^*$.
	The variance contribution due to the intrinsic stochasticity of the motion decreases inversely proportional to the trajectory length, as explained in detail in the SI \cite{flyvbjerg1989error}.
	Therefore, if stochasticity of the motion is the main contribution to the variance, doubling the length of all trajectories of a given population of moving objects leads to ${1-S^*_{\rm{new}}=(1-S^*)/2}$, consequently, $S^*$ increases as ${S^*_{\rm{new}} = S^* + (1-S^*)/2}$.
	In the case of the polystyrene beads for instance, this would mean that for experimental trajectories observed twice as long as in the actual experiment, the estimated covariance ratio would increase from $S^*=0.53$ to roughly $S^*_{\rm{new}}=0.76$, if the spread $1-S^*$ stems mainly from the stochastic process observed over a finite time, i.e. if the influence of other experimental factors on the parameter spread is negligible.

	\section{Discussion \& Conclusions}
	We use the GLE eq. \eqref{eq_gle} to describe active cell motion as well as passive motion of polystyrene beads. The objects'
	motion is described by a set of GLE parameters, which we extract on the single trajectory level.
	The spread of these parameter distributions extracted from the experiments is due to a combination of intrinsic differences among the individual moving 
	objects, the finite trajectory length and noise in the experiment.
	Our simulation approach decomposes how much of the spread actually originates from intrinsic 
	differences between individuals of the observed population and how much stems from the experimental finite observation time.
	On top of that, it allows an estimate of how long experimental trajectories have to be in order to reach a certain accuracy level of the experimentally extracted parameter distributions.
	Our method readily includes experimental factors such as localization noise, discrete time steps and different trajectory lengths.
	We confirm the robustness of our model by applying it to data of passively moving polystyrene beads in water and to synthetic data sets with known origin in the SI.

	Compared to the polystyrene beads that are relatively similar to each other, the cancer cells and algae cells show significant differences within the studied populations.
	For the motion of the breast-cancer cells, most of the observed spread in the mean squared velocity $B$ and in the persistence time $\tau_m$ originates from sizable intrinsic differences on the cell-to-cell level;
	this is noteworthy, as these cells carry the same genetic material.
	Such phenotypic differences were previously observed \cite{snijder2011origins} and are connected to different levels of expressed proteins in the cells \cite{spencer2009non}.
	The observed GLE parameter distribution of algae cells turns out to almost entirely reflect intrinsic cell-to-cell differences, 
	which is part of the reason why the observed GLE parameters could previously be directly 
	used to distinguish different swimming styles of these confined algae cells on the single cell level \cite{klimek_data-driven_2024}.

	The usage of the GLE eq. \eqref{eq_gle} to describe active systems relies on the Gaussianity of the underlying motion process \cite{klimek_data-driven_2024,roland_neq_2023}.
	In cases of non-Gaussian single-cell velocity distributions, one could use filter approaches that render data Gaussian \cite{NetzFilter2024} 
	and then apply the same methodology we describe in this paper.
	Alternatively, for non-Gaussian systems one could use the GLE in a heuristic manner or use previously introduced models 
	to describe the organism motion \cite{bechinger2016active,viswanathan2011physics,dieterich2008anomalous_klages,sadhu2023minimal,heyn2024cell}.
	Since the GLE eq. \eqref{eq_gle} is valid for general observables, our approach can be applied to time-series data characterizing any organism or object of choice - be it active or passive.

	In summary, our approach allows us to disentangle the intrinsic individual-to-individual parameter variance of moving objects from noise and stochastic effects.
	As every distribution of parameters extracted from experiments inevitably is noisy due to finite recording time, we anticipate various applications of our method in infection biology, ecology, medicine and all other fields that rely on the knowledge of the heterogeneity of individual moving objects.

	\section*{Acknowledgement}
		
		We acknowledge the funding by the Deutsche Forschungsgemeinschaft (DFG) through grant CRC 1449 “Dynamic Hydrogels at Biointerfaces”, Project ID 431232613, Project A03.

	\section{Methods, experimental details and cell preparation}

	\subsection{Statistical test}
	In order to compare two $d$-dimensional distributions $\rho^{\rm{exp}}(\vec{z})$ and $\rho^{\rm{sim}}(\vec{z})$ which are determined by a finite number of observations $\vec{z}^{\rm{exp}}_i$ for $\rho^{\rm{exp}}$ and $\vec{z}^{\rm{sim}}_i$ for $\rho^{\rm{sim}}$, we use a so-called permutation test \cite{pesarin2010permutation}. We define the distance measure
	\begin{equation}
		\label{eq_test_statistic}
		T = \sum_{i\leq j}^{d} \left( \rm{Cov}^{\rm{1}}_{ij} - \rm{Cov}^{\rm{2}}_{ij} \right)^2\,,
	\end{equation}
	which measures the difference between two distributions in terms of their covariances $\rm{Cov}^{\rm{1}}$ and $\rm{Cov}^{\rm{2}}$. We denote the value of $T$ of the distributions $\rho^{\rm{exp}}(\vec{z})$ and $\rho^{\rm{sim}}(\vec{z})$ by $T_0$.
	
	The statistical test starts by randomly drawing samples from the joint pool of observations $\vec{z}^{\rm{exp}}_i$ and $\vec{z}^{\rm{sim}}_i$, which leads to the exchanged distributions $\rho^{\rm{exp}}_{\rm{exch}}(\vec{z})$ and $\rho^{\rm{sim}}_{\rm{exch}}(\vec{z})$.
	The exchanged distributions exhibit a new value of $T$ called $T_{\rm{exch}}$.
	This process of sampling exchanged distributions from the joint pool of observations is repeated $n_{\rm{exch}}=10^4$ times, leading to $10^4$ values of $T_{\rm{exch}}$.
	Comparing $T_0$ to $T_{\rm{exch}}$ leads to the probability of rejecting the hypothesis that the two distributions have the same covariance, given that the hypothesis is actually true as ${p_1 = n_{\rm{rej}} / n_{\rm{exch}}}$, with $n_{\rm{rej}}$ being the number of exchanges for which $T_{\rm{exch}} > T_0$ \cite{pesarin2010permutation}.
	We use the acceptance level $p_1^*=0.05$, meaning we accept the hypothesis that the distributions $\rho^{\rm{exp}}(\vec{z})$ and $\rho^{\rm{sim}}(\vec{z})$ are identical in covariance when $p_1\leq p_1^*$.
	
	In practice, we extract $M=100$ distributions $\rho^{\rm{sim}}(\vec{z})$ from simulations and compare each one to the experimental distribution using the statistical test described above.
	This means we simulate a total of $M\times N$ trajectories.
	We define $p_2$ as the ratio of rejected distributions $M_{\rm{rej}}$, where $p_1>p_1^*$, to the total number of simulated distributions $p_2=M_{\rm{rej}}/M$.
	The hypothesis that the covariance from simulation $\rm{Cov}^{\rm{sim}}$ originates from the same intrinsic covariance $\rm{Cov}^{\rm{inp*}}$ as the experimental covariance $\rm{Cov}^{\rm{exp}}$ is accepted as true, if $p_2\leq p_2^*$, where $p_2^*$ is the acceptance level.
	One expects the hypothesis to be rejected in average $p_1^* \times M$ out of $M$ times, if it is true.
	Therefore we set another acceptance level $p_2^*=p_1^*=0.05$.
	Results of $p_2$ depending on $S$ are shown in Figs. \ref{fig_barplot}b-d for different moving objects.

	\subsection{Discretized fit to data including localization noise}\label{sec_fitting}
	Every experimental recording of trajectories inevitably includes noise.
	In the case of cell-center trajectories the finite pixel size, thermal and electronic fluctuations in the camera sensors are just a few of many possible origins of localization noise present in the trajectories.
	Therefore, our fit to the experimental VACF of individual trajectories needs to account for finite time discretization and noise \cite{mitterwallner_non-markovian_2020}.
	Here, we explain this fitting procedure.
	We denote discrete values of a function $f(t)$ as
	$f(i \Delta)=f_i=f^i$ and the discretization time step as $\Delta$.
	Velocities are computed at half time steps as 
	\begin{equation}
		\label{eq_central_diff_vel}
		v_{i+\frac{1}{2}} = \frac{x_{i+1} - x_{i}}{\Delta}\,.
	\end{equation}
	From the velocities the VACF defined by eq. \eqref{eq_def_cvv}
	is computed according to
	\begin{equation}
		\label{eq_vacf_from_data}
		C_{vv}^i=\frac{1}{n-i}\sum_{j=0}^{n-i-1}v_{j+\frac{1}{2}} v_{j+i+\frac{1}{2}}\,,
	\end{equation}
	with $n$ being the the number of trajectory steps.
	In order to account for localization noise, we add Gaussian uncorrelated noise of width $\sigma_{loc}$ at every time step to the trajectory, which leads to the noisy MSD as \cite{mitterwallner_non-markovian_2020}
	\begin{equation}
		\label{eq_msd_noise}
		C_{\textrm{MSD}}^{\rm{noise}}(t)=C_{\textrm{MSD}}^{\rm{theo}}(t)+2(1-\delta_{t0})\sigma_{\textrm{loc}}^2 \,,
	\end{equation}
	where $C_{\textrm{MSD}}^{\rm{theo}}(t)$ is the theoretical expression for the model MSD given by eqs. \eqref{eq_msd_diffusion}, \eqref{eq_msd_prw}, \eqref{eq_msd_final_Res} for the different models eqs. \eqref{eq_kern_osc}, \eqref{eq_kern_prw}, \eqref{eq_gle_diffusion_lim}
	and $\delta_{t0}$ is the Kronecker delta reflecting the uncorrelated nature of the localization noise.
	Since the observed trajectories are sampled with a finite time step $\Delta$, we discretize the relation
	\begin{equation}
		C_{vv}(t) = \frac{1}{2}\frac{d^2}{dt^2}C_{\textrm{MSD}}(t)\,,
		\label{eq_cvv_from_msd}
	\end{equation}
	which leads to
	\begin{align}
		\label{eq_vacf_fit}
		&C_{vv}^{\rm{fit}}(i \Delta) = \nonumber \\ &\frac{C_{\textrm{MSD}}^{\rm{noise}}((i+1)\Delta)-2C_{\textrm{MSD}}^{\rm{noise}}(i \Delta)+C_{\textrm{MSD}}^{\rm{noise}}((i-1)\Delta)}{2\Delta^2} \,.
	\end{align}
	Finally, fits are performed by minimizing the cost function
	\begin{equation}
		\label{eq_cost_function}
		E_{\textrm{cost}}=\sum_{i=0}^{n-1} (C_{vv}^{\rm{exp}}(i \Delta)-C_{vv}^{\rm{fit}}(i \Delta))^2
	\end{equation}
	with scipy's least squares function in python and using eq. \eqref{eq_vacf_fit} to determine the parameters of $C_{vv}^{\rm{fit}}(t)$.
	As the MSD and VACF follow from the GLE 
	\eqref{eq_gle} and the friction kernel in eqs. \eqref{eq_kern_osc},\eqref{eq_kern_prw} or the LE in eq. \eqref{eq_gle_diffusion_lim}, the parameters to be optimized are the kernel parameters, together with the mean squared velocity $B$ and the localization noise width $\sigma_{\rm{loc}}$.
	Resulting fits are shown in Figs. \ref{fig_3}e,h, where the data is fitted up to $3.3\,\rm{min}$ for the cancer cells and up to $0.2\unit{s}$ for the algae cells in order to disregard the noisy part of the VACF at longer times.
	For the polystyrene beads, which perform purely diffusive motion, we use eq. \eqref{eq_msd_noise} to fit the experimental MSD as the VACF is only non-zero for the first two time steps.
	A resulting fit is shown in Fig. \ref{fig_3}a, where the MSD is fitted up to $0.2\unit{s}$.
	
	Before we fit the data on the individual trajectory level, we estimate the order of magnitude of the mean squared velocity $B$ by $B_{\rm{est}} = C_{vv}^0$ for the cancer cells and algae cells.
	For the freely diffusing beads we estimate the diffusion coefficient $D$ by the integral over the VACF, which becomes $D_{\rm{est}} = C_{vv}^0 \frac{\Delta}{2}$.
	Furthermore, for the cancer cells we estimate the order of magnitude of the persistence time $\tau_m$ by $\tau_m^{\rm{est}} = \frac{2}{\Delta \Gamma_0}$.
	We apply boundaries for the optimization of cancer cell parameters: $B$ is constrained between $B_{\rm{est}}/4$ and $4B_{\rm{est}}$, $\tau_m$ between $\tau_m^{\rm{est}}/4$ and $4\tau_m^{\rm{est}}$ and the localization noise width $\sigma_{\rm{loc}}$ between $0.01-1\unit{\mu m}$.
	For the algae cells $B$ is assumed to lie between $B_{\rm{est}}/3$ and $3B_{\rm{est}}$, the delta amplitude $a$ and the oscillation amplitude $b$ are assumed to lie between $0.01\Gamma_0$ and $\Gamma_0$, the memory time $\tau$ is bound between $0.05-5\unit{s}$, the kernel frequency $\Omega$ between $15-250\unit{Hz}$ and the localization noise width $\sigma_{\rm{loc}}$ between $0-0.05\unit{\mu m}$.
	Further, for polystyrene beads we use bounds of $4D_{\rm{est}}$ and $D_{\rm{est}}/4$ for the diffusivity $D$ and the localization noise width $\sigma_{\rm{loc}}$ is constrained between $0.001-0.2\unit{\mu m}$.
	Finally, we discard fit results for which the localization noise width $\sigma_{\rm{loc}}$ is close to the boundaries and likely to be an unphysical fit result, when $\sigma_{\rm{loc}}$ is smaller than $1.5$ times the lower boundary or larger than $0.99$ times the upper boundary.
	This leads to four discarded trajectories out of 135 total trajectories for the breast-cancer cells, no discarded trajectories for the 23 algae cells and 19 discarded trajectories out of 144 for the polystyrene beads.
	For the diffusion of the polystyrene beads we additionally discard diffusion constants above $0.8\unit{\mu m^2/s}$, which is roughly twice the theoretical value of a sphere with radius $r=0.5\unit{\mu m}$ according to the Stokes-Einstein law eq. \eqref{eq_stokes_einstein}, which leads to four additionally discarded trajectories.

	\subsection{MSD expressions}
	
	The diffusion coefficient defined in eq. \eqref{eq_stokes_einstein} dictates the behavior of the MSD for pure diffusion as
	\begin{equation}
		\label{eq_msd_diffusion}
		C_{\rm{MSD}}(t) = 2 Dt\,.
	\end{equation}
	In the case of the PRW the MSD takes the form
	\begin{equation}
		\label{eq_msd_prw}
		C_{\rm{MSD}}(t) = 2 B\tau_m \left( t - \tau_m (1 - e^{-t/\tau_m}) \right)\,,
	\end{equation}
	where the diffusion constant is given by $D=B\tau_m$ and the VACF follows a single exponential decay
	\begin{equation}
		\label{eq_vacf_prw}
		C_{vv}(t) = B e^{-t/\tau_m}\,.
	\end{equation}
	For the oscillating memory kernel eq. \eqref{eq_kern_osc} the MSD takes the form \cite{klimek_data-driven_2024}
	\begin{equation}
		\begin{split}
			&C_{\textrm{MSD}}(t) = \frac{B}{\tau^4} \bigg( \frac{k_1 t}{\omega_1^2 \omega_2^2 \omega_3^2}\\
			&+ \sum_{i=1}^{3}\frac{e^{-\sqrt{-\omega_i^2} t} - 1}{\sqrt{-\omega_i^2} \prod_{j\neq i} (\omega_i^2-\omega_j^2) } \left[ \frac{k_1}{\omega_i^2} + k_2 + k_3 \omega_i^2 \right] \bigg) \,,
		\end{split}
		\label{eq_msd_final_Res}
	\end{equation}
	where the constants $\omega_i$ and $k_i$ have been previously calculated \cite{klimek_data-driven_2024}.

	\subsection{Polystyrene beads}
	
	\textit{Materials:} Fluorescently labeled polystyrene beads (diameter: $1000\unit{nm}$, excitation wavelength: $505\unit{nm}$, emission wavelength: $515\unit{nm}$) were purchased from ThermoFisher Scientific (order number: F8776).
	For single particle tracking experiments, the beads are diluted 1:1000 in deionized water and sonicated for five minutes at room temperature.
	Afterwards, $8\unit{\mu l}$ of bead-containing solution is injected into polydimethylsiloxane (PDMS) microwells, which are formed on glass cover slips as described previously \cite{wallert2020mucin}.
	
	\textit{Imaging:} Single-particle tracking of the fluorescently labeled polystyrene beads is performed using a Nikon Ti-E Eclipse fluorescence microscope (Nikon, Düsseldorf, Germany), which is equipped with focus stabilization, a white light source (Prior Lumen 200; Prior Scientific, Cambridge, UK) and an Andor Zyla 4.2 sCMOS camera (Oxford Instruments, Oxford, UK).
	Fluorescence excitation and emission is controlled using a GFP filter set (Nikon GFP-1828A; EX 482/18, DM 495, BA 520/28).
	The beads are imaged using a 100x Plan-Apo oil immersion objective (numerical aperture: 1.45) and the following setting of the sCMOS camera: $2\times2$ binning, $10\unit{ms}$ exposure time, acquisition rate of 50.41 frames per second at a field of view of $1024\times1024 \unit{pixel^2}$, corresponding to $133.12\times133.12\unit{\mu m^2}$.
	
	\textit{Image analysis:} Single particle tracking analysis is performed as previously described \cite{muller2019mobility}.
	In brief, the beads are detected by a local maximum of the fluorescence intensity and only considered for further analysis, if the corresponding intensity value exceeds a user-defined threshold, which is chosen slightly above the level of readout noise of the sCMOS camera.
	The center position of each detected bead is determined by fitting a two-dimensional Gaussian distribution to its intensity distribution.
	Trajectories are generated by a nearest-neighbor linking scheme involving a distance threshold as described previously \cite{kerkhoff2020analysis}.
	Finally, as convection of the aqueous solution (contained in the PDMS wells) cannot be ruled out even for these relatively small wells, the bead trajectories are corrected for the potential occurrence of convection-based distortions of bead motion.
	Convection affects the displacements of all beads in the same way and there is no further source of motional correlation for these randomly moving beads, except for hydrodynamic interactions, which are weak for the high dilution of beads employed here \cite{von2012auto}.
	The convection is determined (i) by calculating the displacements of all beads between consecutive frames, followed (ii) by calculating the average value of all obtained two-dimensional displacement vectors and finally (iii) by integrating all average displacement vector values from the first to the last frame (whereas the average displacement vector between the first two frames is set to zero).
	This running integral yields an estimate for the convection-based motion of the aqueous solution, which we subtract from the position of the beads to correct the trajectories for convection.
	
	\subsection{Breast-cancer cells}
	
	\textit{Micropatterning:} The human breast-cancer cells of the cell line MDA-MBA-231 in our experiments move on one-dimensional lanes that are produced by coating with fibronectin (FN).
	We transfer fibronectin (FN) (YO Proteins, Ronninge, Sweden) to an imaging dish featuring a polymer coverslip bottom (ibidi, Gräfelfing, Germany) using polydimethylsiloxane (PDMS) stamps with a $15\unit{\mu m}$ wide lane pattern. The microcontact printing protocol, including the fabrication of the PDMS stamps, has been detailed previously \cite{schreiber2016ring}.
	
	\textit{Cell culture:} We culture MDA-MB-231 cells that had been stably transduced with histone-2B mCherry (gift from Timo Betz, University of Göttingen, Germany) in Leibovitz’s \ch{CO_2}-buffered L-15 medium with $2\unit{mM}$ Glutamax (Thermo Fisher Scientific, Waltham, MA, USA) at $37\degC$.
	The growth medium was supplemented by 10\,\% fetal bovine serum (Thermo Fisher) and cells were passaged every 2-3 days using Accutase (Thermo Fisher).
	For experiments, we seeded about 5000 cells per dish. After incubation for 2-3 h, cells adhered to the micropatterns and we exchanged the medium.
	
	\textit{Microscopy:} The microscopy images, see for instance Fig. \ref{fig_0}b, originate from time-lapse imaging on an inverted fluorescence microscope (Nikon Eclipse Ti, Nikon, Tokyo, Japan) outfitted with an XY-motorized stage, Perfect Focus System (Nikon), and a heating chamber (Okolab, Pozzuoli, Italy) maintained at $37\degC$.
	We set up an acquisition protocol to sequentially scan and image fields of view employing the motorized stage, the Perfect Focus System, a 10 CFI Plan Fluor DL objective (Nikon), a CMOS camera (PCO edge 4.2, Excelitas PCO, Kelheim, Germany) and the acquisition software NIS Elements (Nikon).
	Prior to the time-lapse measurement, we obtain epifluorescence images of the FN patterns. Subsequently, we capture phase-contrast images of the cells and epifluorescence images of their nuclei at $20\unit{s}$ intervals for a total of $48\unit{h}$.
	
	\textit{Image analysis:} In order to obtain trajectories from the microscopy data of the fluorescently labeled cell nuclei, we employ MATLAB \cite{Matlabtrack} scripts built upon previous work \cite{schreiber2021adhesion} for image analysis.
	Here, the geometric center of the fluorescently labeled nucleus is used as the cell position. Detection of FN lanes is accomplished through a Hough transformation of the fluorescence signal from labeled FN.
	Tracking of nuclei positions involves setting a threshold after applying a background correction and a band-pass filter to fluorescent images of the nuclei.
	The position of the nuclei is adjusted to ensure that the position aligns with the FN lanes such that nuclei center positions cannot be extracted offside of the FN lane.
	
	\subsection{Algae cells}
	Wild-type Chlamydomonas reinhardtii cells are recorded in two-dimensional confinement between two anti-adhesively coated glass plates with separation of roughly $10\unit{\mu m}$ by high-speed video microscopy (Olympus IX83/IX73) at 500 frames per second with a 40x phase-contrast objective (Olympus, 0.65 NA, Plan N, PH2) connected to a metal oxide semiconductor (CMOS) camera (Phantom Miro C110, Vision Research, pixel size = $5.6\unit{\mu m}$).
	Trajectories of the algae cells are extracted by binarizing the image sequences with appropriate threshold parameters and tracking their 
	centers using standard MATLAB routines \cite{Matlabtrack}.
	More experimental details are given in \cite{mondal_strong_2021,klimek_data-driven_2024}.
	The Chlamydomonas reinhardtii cells were previously shown to exhibit two distinct swimming styles \cite{klimek_data-driven_2024}, only data of cells with synchronous flagella motion is considered in this work.
	The studied algae cells originate from multiple isolated colonies and, hence, are genetically heterogeneous.


\begin{thebibliography}{51}%
		\makeatletter
		\providecommand \@ifxundefined [1]{%
			\@ifx{#1\undefined}
		}%
		\providecommand \@ifnum [1]{%
			\ifnum #1\expandafter \@firstoftwo
			\else \expandafter \@secondoftwo
			\fi
		}%
		\providecommand \@ifx [1]{%
			\ifx #1\expandafter \@firstoftwo
			\else \expandafter \@secondoftwo
			\fi
		}%
		\providecommand \natexlab [1]{#1}%
		\providecommand \enquote  [1]{``#1''}%
		\providecommand \bibnamefont  [1]{#1}%
		\providecommand \bibfnamefont [1]{#1}%
		\providecommand \citenamefont [1]{#1}%
		\providecommand \href@noop [0]{\@secondoftwo}%
		\providecommand \href [0]{\begingroup \@sanitize@url \@href}%
		\providecommand \@href[1]{\@@startlink{#1}\@@href}%
		\providecommand \@@href[1]{\endgroup#1\@@endlink}%
		\providecommand \@sanitize@url [0]{\catcode `\\12\catcode `\$12\catcode
			`\&12\catcode `\#12\catcode `\^12\catcode `\_12\catcode `\%12\relax}%
		\providecommand \@@startlink[1]{}%
		\providecommand \@@endlink[0]{}%
		\providecommand \url  [0]{\begingroup\@sanitize@url \@url }%
		\providecommand \@url [1]{\endgroup\@href {#1}{\urlprefix }}%
		\providecommand \urlprefix  [0]{URL }%
		\providecommand \Eprint [0]{\href }%
		\providecommand \doibase [0]{https://doi.org/}%
		\providecommand \selectlanguage [0]{\@gobble}%
		\providecommand \bibinfo  [0]{\@secondoftwo}%
		\providecommand \bibfield  [0]{\@secondoftwo}%
		\providecommand \translation [1]{[#1]}%
		\providecommand \BibitemOpen [0]{}%
		\providecommand \bibitemStop [0]{}%
		\providecommand \bibitemNoStop [0]{.\EOS\space}%
		\providecommand \EOS [0]{\spacefactor3000\relax}%
		\providecommand \BibitemShut  [1]{\csname bibitem#1\endcsname}%
		\let\auto@bib@innerbib\@empty
		\bibitem [{\citenamefont {Levien}\ \emph {et~al.}(2021)\citenamefont {Levien},
			\citenamefont {Min}, \citenamefont {Kondev},\ and\ \citenamefont
			{Amir}}]{levien2021non}%
		\BibitemOpen
		\bibfield  {author} {\bibinfo {author} {\bibfnamefont {E.}~\bibnamefont
				{Levien}}, \bibinfo {author} {\bibfnamefont {J.}~\bibnamefont {Min}},
			\bibinfo {author} {\bibfnamefont {J.}~\bibnamefont {Kondev}},\ and\ \bibinfo
			{author} {\bibfnamefont {A.}~\bibnamefont {Amir}},\ }\bibfield  {title}
		{\bibinfo {title} {Non-genetic variability in microbial populations: survival
				strategy or nuisance?},\ }\href@noop {} {\bibfield  {journal} {\bibinfo
				{journal} {Reports on progress in physics}\ }\textbf {\bibinfo {volume}
				{84}},\ \bibinfo {pages} {116601} (\bibinfo {year} {2021})}\BibitemShut
		{NoStop}%
		\bibitem [{\citenamefont {Conner}\ and\ \citenamefont
			{White}(1999)}]{conner1999effects}%
		\BibitemOpen
		\bibfield  {author} {\bibinfo {author} {\bibfnamefont {M.~M.}\ \bibnamefont
				{Conner}}\ and\ \bibinfo {author} {\bibfnamefont {G.~C.}\ \bibnamefont
				{White}},\ }\bibfield  {title} {\bibinfo {title} {Effects of individual
				heterogeneity in estimating the persistence of small populations},\
		}\href@noop {} {\bibfield  {journal} {\bibinfo  {journal} {Natural Resource
					Modeling}\ }\textbf {\bibinfo {volume} {12}},\ \bibinfo {pages} {109}
			(\bibinfo {year} {1999})}\BibitemShut {NoStop}%
		\bibitem [{\citenamefont {Fox}\ \emph {et~al.}(2006)\citenamefont {Fox},
			\citenamefont {Kendall}, \citenamefont {Fitzpatrick},\ and\ \citenamefont
			{Woolfenden}}]{fox2006consequences}%
		\BibitemOpen
		\bibfield  {author} {\bibinfo {author} {\bibfnamefont {G.~A.}\ \bibnamefont
				{Fox}}, \bibinfo {author} {\bibfnamefont {B.~E.}\ \bibnamefont {Kendall}},
			\bibinfo {author} {\bibfnamefont {J.~W.}\ \bibnamefont {Fitzpatrick}},\ and\
			\bibinfo {author} {\bibfnamefont {G.~E.}\ \bibnamefont {Woolfenden}},\
		}\bibfield  {title} {\bibinfo {title} {Consequences of heterogeneity in
				survival probability in a population of florida scrub-jays},\ }\href@noop {}
		{\bibfield  {journal} {\bibinfo  {journal} {Journal of Animal Ecology}\
			}\textbf {\bibinfo {volume} {75}},\ \bibinfo {pages} {921} (\bibinfo {year}
			{2006})}\BibitemShut {NoStop}%
		\bibitem [{\citenamefont {Avraham}\ \emph {et~al.}(2015)\citenamefont
			{Avraham}, \citenamefont {Haseley}, \citenamefont {Brown}, \citenamefont
			{Penaranda}, \citenamefont {Jijon}, \citenamefont {Trombetta}, \citenamefont
			{Satija}, \citenamefont {Shalek}, \citenamefont {Xavier}, \citenamefont
			{Regev} \emph {et~al.}}]{avraham2015pathogen}%
		\BibitemOpen
		\bibfield  {author} {\bibinfo {author} {\bibfnamefont {R.}~\bibnamefont
				{Avraham}}, \bibinfo {author} {\bibfnamefont {N.}~\bibnamefont {Haseley}},
			\bibinfo {author} {\bibfnamefont {D.}~\bibnamefont {Brown}}, \bibinfo
			{author} {\bibfnamefont {C.}~\bibnamefont {Penaranda}}, \bibinfo {author}
			{\bibfnamefont {H.~B.}\ \bibnamefont {Jijon}}, \bibinfo {author}
			{\bibfnamefont {J.~J.}\ \bibnamefont {Trombetta}}, \bibinfo {author}
			{\bibfnamefont {R.}~\bibnamefont {Satija}}, \bibinfo {author} {\bibfnamefont
				{A.~K.}\ \bibnamefont {Shalek}}, \bibinfo {author} {\bibfnamefont {R.~J.}\
				\bibnamefont {Xavier}}, \bibinfo {author} {\bibfnamefont {A.}~\bibnamefont
				{Regev}}, \emph {et~al.},\ }\bibfield  {title} {\bibinfo {title} {Pathogen
				cell-to-cell variability drives heterogeneity in host immune responses},\
		}\href@noop {} {\bibfield  {journal} {\bibinfo  {journal} {Cell}\ }\textbf
			{\bibinfo {volume} {162}},\ \bibinfo {pages} {1309} (\bibinfo {year}
			{2015})}\BibitemShut {NoStop}%
		\bibitem [{\citenamefont {Fleming-Davies}\ \emph {et~al.}(2015)\citenamefont
			{Fleming-Davies}, \citenamefont {Dukic}, \citenamefont {Andreasen},\ and\
			\citenamefont {Dwyer}}]{fleming2015effects}%
		\BibitemOpen
		\bibfield  {author} {\bibinfo {author} {\bibfnamefont {A.~E.}\ \bibnamefont
				{Fleming-Davies}}, \bibinfo {author} {\bibfnamefont {V.}~\bibnamefont
				{Dukic}}, \bibinfo {author} {\bibfnamefont {V.}~\bibnamefont {Andreasen}},\
			and\ \bibinfo {author} {\bibfnamefont {G.}~\bibnamefont {Dwyer}},\ }\bibfield
		{title} {\bibinfo {title} {Effects of host heterogeneity on pathogen
				diversity and evolution},\ }\href@noop {} {\bibfield  {journal} {\bibinfo
				{journal} {Ecology letters}\ }\textbf {\bibinfo {volume} {18}},\ \bibinfo
			{pages} {1252} (\bibinfo {year} {2015})}\BibitemShut {NoStop}%
		\bibitem [{\citenamefont {Asadullah}\ \emph {et~al.}(2021)\citenamefont
			{Asadullah}, \citenamefont {Kumar}, \citenamefont {Saxena}, \citenamefont
			{Sarkar}, \citenamefont {Barai},\ and\ \citenamefont
			{Sen}}]{asadullah2021combined}%
		\BibitemOpen
		\bibfield  {author} {\bibinfo {author} {\bibnamefont {Asadullah}}, \bibinfo
			{author} {\bibfnamefont {S.}~\bibnamefont {Kumar}}, \bibinfo {author}
			{\bibfnamefont {N.}~\bibnamefont {Saxena}}, \bibinfo {author} {\bibfnamefont
				{M.}~\bibnamefont {Sarkar}}, \bibinfo {author} {\bibfnamefont
				{A.}~\bibnamefont {Barai}},\ and\ \bibinfo {author} {\bibfnamefont
				{S.}~\bibnamefont {Sen}},\ }\bibfield  {title} {\bibinfo {title} {Combined
				heterogeneity in cell size and deformability promotes cancer invasiveness},\
		}\href@noop {} {\bibfield  {journal} {\bibinfo  {journal} {Journal of cell
					science}\ }\textbf {\bibinfo {volume} {134}},\ \bibinfo {pages} {jcs250225}
			(\bibinfo {year} {2021})}\BibitemShut {NoStop}%
		\bibitem [{\citenamefont {Aird}(2003)}]{aird2003endothelial}%
		\BibitemOpen
		\bibfield  {author} {\bibinfo {author} {\bibfnamefont {W.~C.}\ \bibnamefont
				{Aird}},\ }\bibfield  {title} {\bibinfo {title} {Endothelial cell
				heterogeneity},\ }\href@noop {} {\bibfield  {journal} {\bibinfo  {journal}
				{Critical care medicine}\ }\textbf {\bibinfo {volume} {31}},\ \bibinfo
			{pages} {S221} (\bibinfo {year} {2003})}\BibitemShut {NoStop}%
		\bibitem [{\citenamefont {Rigler}\ \emph {et~al.}(1999)\citenamefont {Rigler},
			\citenamefont {Pramanik}, \citenamefont {Jonasson}, \citenamefont {Kratz},
			\citenamefont {Jansson}, \citenamefont {Nygren}, \citenamefont {St{\aa}hl},
			\citenamefont {Ekberg}, \citenamefont {Johansson}, \citenamefont {Uhlen}
			\emph {et~al.}}]{rigler1999specific}%
		\BibitemOpen
		\bibfield  {author} {\bibinfo {author} {\bibfnamefont {R.}~\bibnamefont
				{Rigler}}, \bibinfo {author} {\bibfnamefont {A.}~\bibnamefont {Pramanik}},
			\bibinfo {author} {\bibfnamefont {P.}~\bibnamefont {Jonasson}}, \bibinfo
			{author} {\bibfnamefont {G.}~\bibnamefont {Kratz}}, \bibinfo {author}
			{\bibfnamefont {O.}~\bibnamefont {Jansson}}, \bibinfo {author} {\bibfnamefont
				{P.-{\AA}.}\ \bibnamefont {Nygren}}, \bibinfo {author} {\bibfnamefont
				{S.}~\bibnamefont {St{\aa}hl}}, \bibinfo {author} {\bibfnamefont
				{K.}~\bibnamefont {Ekberg}}, \bibinfo {author} {\bibfnamefont {B.-L.}\
				\bibnamefont {Johansson}}, \bibinfo {author} {\bibfnamefont {S.}~\bibnamefont
				{Uhlen}}, \emph {et~al.},\ }\bibfield  {title} {\bibinfo {title} {Specific
				binding of proinsulin c-peptide to human cell membranes},\ }\href@noop {}
		{\bibfield  {journal} {\bibinfo  {journal} {Proceedings of the National
					Academy of Sciences}\ }\textbf {\bibinfo {volume} {96}},\ \bibinfo {pages}
			{13318} (\bibinfo {year} {1999})}\BibitemShut {NoStop}%
		\bibitem [{\citenamefont {Jerison}\ and\ \citenamefont
			{Quake}(2020)}]{jerison2020heterogeneous}%
		\BibitemOpen
		\bibfield  {author} {\bibinfo {author} {\bibfnamefont {E.~R.}\ \bibnamefont
				{Jerison}}\ and\ \bibinfo {author} {\bibfnamefont {S.~R.}\ \bibnamefont
				{Quake}},\ }\bibfield  {title} {\bibinfo {title} {Heterogeneous t cell
				motility behaviors emerge from a coupling between speed and turning in
				vivo},\ }\href@noop {} {\bibfield  {journal} {\bibinfo  {journal} {Elife}\
			}\textbf {\bibinfo {volume} {9}},\ \bibinfo {pages} {e53933} (\bibinfo {year}
			{2020})}\BibitemShut {NoStop}%
		\bibitem [{\citenamefont {Studenski}\ \emph {et~al.}(2011)\citenamefont
			{Studenski}, \citenamefont {Perera}, \citenamefont {Patel}, \citenamefont
			{Rosano}, \citenamefont {Faulkner}, \citenamefont {Inzitari}, \citenamefont
			{Brach}, \citenamefont {Chandler}, \citenamefont {Cawthon}, \citenamefont
			{Connor} \emph {et~al.}}]{studenski2011gait}%
		\BibitemOpen
		\bibfield  {author} {\bibinfo {author} {\bibfnamefont {S.}~\bibnamefont
				{Studenski}}, \bibinfo {author} {\bibfnamefont {S.}~\bibnamefont {Perera}},
			\bibinfo {author} {\bibfnamefont {K.}~\bibnamefont {Patel}}, \bibinfo
			{author} {\bibfnamefont {C.}~\bibnamefont {Rosano}}, \bibinfo {author}
			{\bibfnamefont {K.}~\bibnamefont {Faulkner}}, \bibinfo {author}
			{\bibfnamefont {M.}~\bibnamefont {Inzitari}}, \bibinfo {author}
			{\bibfnamefont {J.}~\bibnamefont {Brach}}, \bibinfo {author} {\bibfnamefont
				{J.}~\bibnamefont {Chandler}}, \bibinfo {author} {\bibfnamefont
				{P.}~\bibnamefont {Cawthon}}, \bibinfo {author} {\bibfnamefont {E.~B.}\
				\bibnamefont {Connor}}, \emph {et~al.},\ }\bibfield  {title} {\bibinfo
			{title} {Gait speed and survival in older adults},\ }\href@noop {} {\bibfield
			{journal} {\bibinfo  {journal} {Jama}\ }\textbf {\bibinfo {volume} {305}},\
			\bibinfo {pages} {50} (\bibinfo {year} {2011})}\BibitemShut {NoStop}%
		\bibitem [{\citenamefont {Maiuri}\ \emph {et~al.}(2015)\citenamefont {Maiuri},
			\citenamefont {Rupprecht}, \citenamefont {Wieser}, \citenamefont {Ruprecht},
			\citenamefont {B{\'e}nichou}, \citenamefont {Carpi}, \citenamefont {Coppey},
			\citenamefont {De~Beco}, \citenamefont {Gov}, \citenamefont {Heisenberg}
			\emph {et~al.}}]{maiuri2015actin}%
		\BibitemOpen
		\bibfield  {author} {\bibinfo {author} {\bibfnamefont {P.}~\bibnamefont
				{Maiuri}}, \bibinfo {author} {\bibfnamefont {J.-F.}\ \bibnamefont
				{Rupprecht}}, \bibinfo {author} {\bibfnamefont {S.}~\bibnamefont {Wieser}},
			\bibinfo {author} {\bibfnamefont {V.}~\bibnamefont {Ruprecht}}, \bibinfo
			{author} {\bibfnamefont {O.}~\bibnamefont {B{\'e}nichou}}, \bibinfo {author}
			{\bibfnamefont {N.}~\bibnamefont {Carpi}}, \bibinfo {author} {\bibfnamefont
				{M.}~\bibnamefont {Coppey}}, \bibinfo {author} {\bibfnamefont
				{S.}~\bibnamefont {De~Beco}}, \bibinfo {author} {\bibfnamefont
				{N.}~\bibnamefont {Gov}}, \bibinfo {author} {\bibfnamefont {C.-P.}\
				\bibnamefont {Heisenberg}}, \emph {et~al.},\ }\bibfield  {title} {\bibinfo
			{title} {Actin flows mediate a universal coupling between cell speed and cell
				persistence},\ }\href@noop {} {\bibfield  {journal} {\bibinfo  {journal}
				{Cell}\ }\textbf {\bibinfo {volume} {161}},\ \bibinfo {pages} {374} (\bibinfo
			{year} {2015})}\BibitemShut {NoStop}%
		\bibitem [{\citenamefont {Klimek}\ \emph {et~al.}(2024)\citenamefont {Klimek},
			\citenamefont {Mondal}, \citenamefont {Block}, \citenamefont {Sharma},\ and\
			\citenamefont {Netz}}]{klimek_data-driven_2024}%
		\BibitemOpen
		\bibfield  {author} {\bibinfo {author} {\bibfnamefont {A.}~\bibnamefont
				{Klimek}}, \bibinfo {author} {\bibfnamefont {D.}~\bibnamefont {Mondal}},
			\bibinfo {author} {\bibfnamefont {S.}~\bibnamefont {Block}}, \bibinfo
			{author} {\bibfnamefont {P.}~\bibnamefont {Sharma}},\ and\ \bibinfo {author}
			{\bibfnamefont {R.~R.}\ \bibnamefont {Netz}},\ }\bibfield  {title} {\bibinfo
			{title} {Data-driven classification of individual cells by their
				non-{Markovian} motion},\ }\bibfield  {journal} {\bibinfo  {journal}
			{Biophysical Journal}\ }\href {https://doi.org/10.1016/j.bpj.2024.03.023}
		{10.1016/j.bpj.2024.03.023} (\bibinfo {year} {2024})\BibitemShut {NoStop}%
		\bibitem [{\citenamefont {Br{\"u}nig}\ \emph
			{et~al.}(2022{\natexlab{a}})\citenamefont {Br{\"u}nig}, \citenamefont
			{Daldrop},\ and\ \citenamefont {Netz}}]{Florian2022a}%
		\BibitemOpen
		\bibfield  {author} {\bibinfo {author} {\bibfnamefont {F.~N.}\ \bibnamefont
				{Br{\"u}nig}}, \bibinfo {author} {\bibfnamefont {J.~O.}\ \bibnamefont
				{Daldrop}},\ and\ \bibinfo {author} {\bibfnamefont {R.~R.}\ \bibnamefont
				{Netz}},\ }\bibfield  {title} {\bibinfo {title} {Pair-reaction dynamics in
				water: competition of memory, potential shape, and inertial effects},\
		}\href@noop {} {\bibfield  {journal} {\bibinfo  {journal} {The Journal of
					Physical Chemistry B}\ }\textbf {\bibinfo {volume} {126}},\ \bibinfo {pages}
			{10295} (\bibinfo {year} {2022}{\natexlab{a}})}\BibitemShut {NoStop}%
		\bibitem [{\citenamefont {Br{\"u}nig}\ \emph
			{et~al.}(2022{\natexlab{b}})\citenamefont {Br{\"u}nig}, \citenamefont
			{Geburtig}, \citenamefont {Canal}, \citenamefont {Kappler},\ and\
			\citenamefont {Netz}}]{Florian2022b}%
		\BibitemOpen
		\bibfield  {author} {\bibinfo {author} {\bibfnamefont {F.~N.}\ \bibnamefont
				{Br{\"u}nig}}, \bibinfo {author} {\bibfnamefont {O.}~\bibnamefont
				{Geburtig}}, \bibinfo {author} {\bibfnamefont {A.~v.}\ \bibnamefont {Canal}},
			\bibinfo {author} {\bibfnamefont {J.}~\bibnamefont {Kappler}},\ and\ \bibinfo
			{author} {\bibfnamefont {R.~R.}\ \bibnamefont {Netz}},\ }\bibfield  {title}
		{\bibinfo {title} {Time-dependent friction effects on vibrational infrared
				frequencies and line shapes of liquid water},\ }\href@noop {} {\bibfield
			{journal} {\bibinfo  {journal} {The Journal of Physical Chemistry B}\
			}\textbf {\bibinfo {volume} {126}},\ \bibinfo {pages} {1579} (\bibinfo {year}
			{2022}{\natexlab{b}})}\BibitemShut {NoStop}%
		\bibitem [{\citenamefont {Dalton}\ \emph {et~al.}(2024)\citenamefont {Dalton},
			\citenamefont {Kiefer},\ and\ \citenamefont {Netz}}]{dalton2024role}%
		\BibitemOpen
		\bibfield  {author} {\bibinfo {author} {\bibfnamefont {B.~A.}\ \bibnamefont
				{Dalton}}, \bibinfo {author} {\bibfnamefont {H.}~\bibnamefont {Kiefer}},\
			and\ \bibinfo {author} {\bibfnamefont {R.~R.}\ \bibnamefont {Netz}},\
		}\bibfield  {title} {\bibinfo {title} {The role of memory-dependent friction
				and solvent viscosity in isomerization kinetics in viscogenic media},\
		}\href@noop {} {\bibfield  {journal} {\bibinfo  {journal} {Nature
					Communications}\ }\textbf {\bibinfo {volume} {15}},\ \bibinfo {pages} {3761}
			(\bibinfo {year} {2024})}\BibitemShut {NoStop}%
		\bibitem [{\citenamefont {Mitterwallner}\ \emph
			{et~al.}(2020{\natexlab{a}})\citenamefont {Mitterwallner}, \citenamefont
			{Schreiber}, \citenamefont {Daldrop}, \citenamefont {R{\"a}dler},\ and\
			\citenamefont {Netz}}]{mitterwallner_non-markovian_2020}%
		\BibitemOpen
		\bibfield  {author} {\bibinfo {author} {\bibfnamefont {B.~G.}\ \bibnamefont
				{Mitterwallner}}, \bibinfo {author} {\bibfnamefont {C.}~\bibnamefont
				{Schreiber}}, \bibinfo {author} {\bibfnamefont {J.~O.}\ \bibnamefont
				{Daldrop}}, \bibinfo {author} {\bibfnamefont {J.~O.}\ \bibnamefont
				{Rädler}},\ and\ \bibinfo {author} {\bibfnamefont {R.~R.}\ \bibnamefont
				{Netz}},\ }\bibfield  {title} {\bibinfo {title} {Non-{Markovian} data-driven
				modeling of single-cell motility},\ }\href
		{https://doi.org/10.1103/PhysRevE.101.032408} {\bibfield  {journal} {\bibinfo
				{journal} {Phys. Rev. E}\ }\textbf {\bibinfo {volume} {101}},\ \bibinfo
			{pages} {032408} (\bibinfo {year} {2020}{\natexlab{a}})},\ \bibinfo {note}
		{publisher: American Physical Society}\BibitemShut {NoStop}%
		\bibitem [{\citenamefont {Ayaz}\ \emph {et~al.}(2021)\citenamefont {Ayaz},
			\citenamefont {Tepper}, \citenamefont {Br{\"u}nig}, \citenamefont {Kappler},
			\citenamefont {Daldrop},\ and\ \citenamefont {Netz}}]{ayaz2021non}%
		\BibitemOpen
		\bibfield  {author} {\bibinfo {author} {\bibfnamefont {C.}~\bibnamefont
				{Ayaz}}, \bibinfo {author} {\bibfnamefont {L.}~\bibnamefont {Tepper}},
			\bibinfo {author} {\bibfnamefont {F.~N.}\ \bibnamefont {Br{\"u}nig}},
			\bibinfo {author} {\bibfnamefont {J.}~\bibnamefont {Kappler}}, \bibinfo
			{author} {\bibfnamefont {J.~O.}\ \bibnamefont {Daldrop}},\ and\ \bibinfo
			{author} {\bibfnamefont {R.~R.}\ \bibnamefont {Netz}},\ }\bibfield  {title}
		{\bibinfo {title} {Non-markovian modeling of protein folding},\ }\href@noop
		{} {\bibfield  {journal} {\bibinfo  {journal} {Proceedings of the National
					Academy of Sciences}\ }\textbf {\bibinfo {volume} {118}},\ \bibinfo {pages}
			{e2023856118} (\bibinfo {year} {2021})}\BibitemShut {NoStop}%
		\bibitem [{\citenamefont {Dalton}\ \emph {et~al.}(2023)\citenamefont {Dalton},
			\citenamefont {Ayaz}, \citenamefont {Kiefer}, \citenamefont {Klimek},
			\citenamefont {Tepper},\ and\ \citenamefont {Netz}}]{dalton2023fast}%
		\BibitemOpen
		\bibfield  {author} {\bibinfo {author} {\bibfnamefont {B.~A.}\ \bibnamefont
				{Dalton}}, \bibinfo {author} {\bibfnamefont {C.}~\bibnamefont {Ayaz}},
			\bibinfo {author} {\bibfnamefont {H.}~\bibnamefont {Kiefer}}, \bibinfo
			{author} {\bibfnamefont {A.}~\bibnamefont {Klimek}}, \bibinfo {author}
			{\bibfnamefont {L.}~\bibnamefont {Tepper}},\ and\ \bibinfo {author}
			{\bibfnamefont {R.~R.}\ \bibnamefont {Netz}},\ }\bibfield  {title} {\bibinfo
			{title} {Fast protein folding is governed by memory-dependent friction},\
		}\href@noop {} {\bibfield  {journal} {\bibinfo  {journal} {Proceedings of the
					National Academy of Sciences}\ }\textbf {\bibinfo {volume} {120}},\ \bibinfo
			{pages} {e2220068120} (\bibinfo {year} {2023})}\BibitemShut {NoStop}%
		\bibitem [{\citenamefont {Netz}(2024{\natexlab{a}})}]{roland_neq_2023}%
		\BibitemOpen
		\bibfield  {author} {\bibinfo {author} {\bibfnamefont {R.~R.}\ \bibnamefont
				{Netz}},\ }\bibfield  {title} {\bibinfo {title} {Derivation of the
				nonequilibrium generalized langevin equation from a time-dependent many-body
				hamiltonian},\ }\href@noop {} {\bibfield  {journal} {\bibinfo  {journal}
				{Physical Review E}\ }\textbf {\bibinfo {volume} {110}},\ \bibinfo {pages}
			{014123} (\bibinfo {year} {2024}{\natexlab{a}})}\BibitemShut {NoStop}%
		\bibitem [{\citenamefont {Martin}\ \emph {et~al.}(2021)\citenamefont {Martin},
			\citenamefont {O'Byrne}, \citenamefont {Cates}, \citenamefont {Fodor},
			\citenamefont {Nardini}, \citenamefont {Tailleur},\ and\ \citenamefont
			{van~Wijland}}]{martin2021statistical_OU}%
		\BibitemOpen
		\bibfield  {author} {\bibinfo {author} {\bibfnamefont {D.}~\bibnamefont
				{Martin}}, \bibinfo {author} {\bibfnamefont {J.}~\bibnamefont {O'Byrne}},
			\bibinfo {author} {\bibfnamefont {M.~E.}\ \bibnamefont {Cates}}, \bibinfo
			{author} {\bibfnamefont {{\'E}.}~\bibnamefont {Fodor}}, \bibinfo {author}
			{\bibfnamefont {C.}~\bibnamefont {Nardini}}, \bibinfo {author} {\bibfnamefont
				{J.}~\bibnamefont {Tailleur}},\ and\ \bibinfo {author} {\bibfnamefont
				{F.}~\bibnamefont {Van~Wijland}},\ }\bibfield  {title} {\bibinfo {title}
			{Statistical mechanics of active ornstein-uhlenbeck particles},\ }\href@noop
		{} {\bibfield  {journal} {\bibinfo  {journal} {Physical Review E}\ }\textbf
			{\bibinfo {volume} {103}},\ \bibinfo {pages} {032607} (\bibinfo {year}
			{2021})}\BibitemShut {NoStop}%
		\bibitem [{\citenamefont {Mitterwallner}\ \emph
			{et~al.}(2020{\natexlab{b}})\citenamefont {Mitterwallner}, \citenamefont
			{Lavacchi},\ and\ \citenamefont {Netz}}]{mitterwallner2020negative}%
		\BibitemOpen
		\bibfield  {author} {\bibinfo {author} {\bibfnamefont {B.~G.}\ \bibnamefont
				{Mitterwallner}}, \bibinfo {author} {\bibfnamefont {L.}~\bibnamefont
				{Lavacchi}},\ and\ \bibinfo {author} {\bibfnamefont {R.~R.}\ \bibnamefont
				{Netz}},\ }\bibfield  {title} {\bibinfo {title} {Negative friction memory
				induces persistent motion},\ }\href@noop {} {\bibfield  {journal} {\bibinfo
				{journal} {The European Physical Journal E}\ }\textbf {\bibinfo {volume}
				{43}},\ \bibinfo {pages} {1} (\bibinfo {year}
			{2020}{\natexlab{b}})}\BibitemShut {NoStop}%
		\bibitem [{\citenamefont {Viswanathan}\ \emph {et~al.}(2011)\citenamefont
			{Viswanathan}, \citenamefont {Da~Luz}, \citenamefont {Raposo},\ and\
			\citenamefont {Stanley}}]{viswanathan2011physics}%
		\BibitemOpen
		\bibfield  {author} {\bibinfo {author} {\bibfnamefont {G.~M.}\ \bibnamefont
				{Viswanathan}}, \bibinfo {author} {\bibfnamefont {M.~G.}\ \bibnamefont
				{Da~Luz}}, \bibinfo {author} {\bibfnamefont {E.~P.}\ \bibnamefont {Raposo}},\
			and\ \bibinfo {author} {\bibfnamefont {H.~E.}\ \bibnamefont {Stanley}},\
		}\href@noop {} {\emph {\bibinfo {title} {The physics of foraging: an
					introduction to random searches and biological encounters}}}\ (\bibinfo
		{publisher} {Cambridge University Press},\ \bibinfo {year}
		{2011})\BibitemShut {NoStop}%
		\bibitem [{\citenamefont {Moon}(1996)}]{moon1996expectation}%
		\BibitemOpen
		\bibfield  {author} {\bibinfo {author} {\bibfnamefont {T.~K.}\ \bibnamefont
				{Moon}},\ }\bibfield  {title} {\bibinfo {title} {The expectation-maximization
				algorithm},\ }\href@noop {} {\bibfield  {journal} {\bibinfo  {journal} {IEEE
					Signal processing magazine}\ }\textbf {\bibinfo {volume} {13}},\ \bibinfo
			{pages} {47} (\bibinfo {year} {1996})}\BibitemShut {NoStop}%
		\bibitem [{\citenamefont {Law}\ \emph {et~al.}(2007)\citenamefont {Law},
			\citenamefont {Kelton},\ and\ \citenamefont {Kelton}}]{law2007simulation}%
		\BibitemOpen
		\bibfield  {author} {\bibinfo {author} {\bibfnamefont {A.~M.}\ \bibnamefont
				{Law}}, \bibinfo {author} {\bibfnamefont {W.~D.}\ \bibnamefont {Kelton}},\
			and\ \bibinfo {author} {\bibfnamefont {W.~D.}\ \bibnamefont {Kelton}},\
		}\href@noop {} {\emph {\bibinfo {title} {Simulation modeling and
					analysis}}},\ Vol.~\bibinfo {volume} {3}\ (\bibinfo  {publisher} {Mcgraw-hill
			New York},\ \bibinfo {year} {2007})\BibitemShut {NoStop}%
		\bibitem [{\citenamefont {Barton}\ and\ \citenamefont
			{Schruben}(2001)}]{barton2001resampling}%
		\BibitemOpen
		\bibfield  {author} {\bibinfo {author} {\bibfnamefont {R.~R.}\ \bibnamefont
				{Barton}}\ and\ \bibinfo {author} {\bibfnamefont {L.~W.}\ \bibnamefont
				{Schruben}},\ }\bibfield  {title} {\bibinfo {title} {Resampling methods for
				input modeling},\ }in\ \href@noop {} {\emph {\bibinfo {booktitle} {Proceeding
					of the 2001 winter simulation conference (Cat. No. 01CH37304)}}},\
		Vol.~\bibinfo {volume} {1}\ (\bibinfo {organization} {IEEE},\ \bibinfo {year}
		{2001})\ pp.\ \bibinfo {pages} {372--378}\BibitemShut {NoStop}%
		\bibitem [{\citenamefont {Massada}\ and\ \citenamefont
			{Carmel}(2008)}]{massada2008incorporating}%
		\BibitemOpen
		\bibfield  {author} {\bibinfo {author} {\bibfnamefont {A.~B.}\ \bibnamefont
				{Massada}}\ and\ \bibinfo {author} {\bibfnamefont {Y.}~\bibnamefont
				{Carmel}},\ }\bibfield  {title} {\bibinfo {title} {Incorporating output
				variance in local sensitivity analysis for stochastic models},\ }\href@noop
		{} {\bibfield  {journal} {\bibinfo  {journal} {ecological modelling}\
			}\textbf {\bibinfo {volume} {213}},\ \bibinfo {pages} {463} (\bibinfo {year}
			{2008})}\BibitemShut {NoStop}%
		\bibitem [{\citenamefont {Schreiber}\ \emph {et~al.}(2021)\citenamefont
			{Schreiber}, \citenamefont {Amiri}, \citenamefont {Heyn}, \citenamefont
			{R{\"a}dler},\ and\ \citenamefont {Falcke}}]{schreiber2021adhesion}%
		\BibitemOpen
		\bibfield  {author} {\bibinfo {author} {\bibfnamefont {C.}~\bibnamefont
				{Schreiber}}, \bibinfo {author} {\bibfnamefont {B.}~\bibnamefont {Amiri}},
			\bibinfo {author} {\bibfnamefont {J.~C.}\ \bibnamefont {Heyn}}, \bibinfo
			{author} {\bibfnamefont {J.~O.}\ \bibnamefont {R{\"a}dler}},\ and\ \bibinfo
			{author} {\bibfnamefont {M.}~\bibnamefont {Falcke}},\ }\bibfield  {title}
		{\bibinfo {title} {On the adhesion--velocity relation and length adaptation
				of motile cells on stepped fibronectin lanes},\ }\href@noop {} {\bibfield
			{journal} {\bibinfo  {journal} {Proceedings of the National Academy of
					Sciences}\ }\textbf {\bibinfo {volume} {118}},\ \bibinfo {pages}
			{e2009959118} (\bibinfo {year} {2021})}\BibitemShut {NoStop}%
		\bibitem [{\citenamefont {Amiri}\ \emph {et~al.}(2023)\citenamefont {Amiri},
			\citenamefont {Heyn}, \citenamefont {Schreiber}, \citenamefont {R{\"a}dler},\
			and\ \citenamefont {Falcke}}]{amiri2023multistability}%
		\BibitemOpen
		\bibfield  {author} {\bibinfo {author} {\bibfnamefont {B.}~\bibnamefont
				{Amiri}}, \bibinfo {author} {\bibfnamefont {J.~C.}\ \bibnamefont {Heyn}},
			\bibinfo {author} {\bibfnamefont {C.}~\bibnamefont {Schreiber}}, \bibinfo
			{author} {\bibfnamefont {J.~O.}\ \bibnamefont {R{\"a}dler}},\ and\ \bibinfo
			{author} {\bibfnamefont {M.}~\bibnamefont {Falcke}},\ }\bibfield  {title}
		{\bibinfo {title} {On multistability and constitutive relations of cell
				motion on fibronectin lanes},\ }\href@noop {} {\bibfield  {journal} {\bibinfo
				{journal} {Biophysical Journal}\ }\textbf {\bibinfo {volume} {122}},\
			\bibinfo {pages} {753} (\bibinfo {year} {2023})}\BibitemShut {NoStop}%
		\bibitem [{\citenamefont {Mondal}\ \emph {et~al.}(2021)\citenamefont {Mondal},
			\citenamefont {Prabhune}, \citenamefont {Ramaswamy},\ and\ \citenamefont
			{Sharma}}]{mondal_strong_2021}%
		\BibitemOpen
		\bibfield  {author} {\bibinfo {author} {\bibfnamefont {D.}~\bibnamefont
				{Mondal}}, \bibinfo {author} {\bibfnamefont {A.~G.}\ \bibnamefont
				{Prabhune}}, \bibinfo {author} {\bibfnamefont {S.}~\bibnamefont
				{Ramaswamy}},\ and\ \bibinfo {author} {\bibfnamefont {P.}~\bibnamefont
				{Sharma}},\ }\bibfield  {title} {\bibinfo {title} {Strong confinement of
				active microalgae leads to inversion of vortex flow and enhanced mixing},\
		}\href {https://doi.org/10.7554/eLife.67663} {\bibfield  {journal} {\bibinfo
				{journal} {eLife}\ }\textbf {\bibinfo {volume} {10}},\ \bibinfo {pages}
			{e67663} (\bibinfo {year} {2021})},\ \bibinfo {note} {publisher: eLife
			Sciences Publications, Ltd}\BibitemShut {NoStop}%
		\bibitem [{\citenamefont {Mori}(1965)}]{mori_transport_1965}%
		\BibitemOpen
		\bibfield  {author} {\bibinfo {author} {\bibfnamefont {H.}~\bibnamefont
				{Mori}},\ }\bibfield  {title} {\bibinfo {title} {Transport, {Collective}
				{Motion}, and {Brownian} {Motion}},\ }\href
		{https://doi.org/10.1143/PTP.33.423} {\bibfield  {journal} {\bibinfo
				{journal} {Prog. Theor. Phys.}\ }\textbf {\bibinfo {volume} {33}},\ \bibinfo
			{pages} {423} (\bibinfo {year} {1965})}\BibitemShut {NoStop}%
		\bibitem [{\citenamefont {Zwanzig}(1961)}]{zwanzig1961memory}%
		\BibitemOpen
		\bibfield  {author} {\bibinfo {author} {\bibfnamefont {R.}~\bibnamefont
				{Zwanzig}},\ }\bibfield  {title} {\bibinfo {title} {Memory effects in
				irreversible thermodynamics},\ }\href@noop {} {\bibfield  {journal} {\bibinfo
				{journal} {Physical Review}\ }\textbf {\bibinfo {volume} {124}},\ \bibinfo
			{pages} {983} (\bibinfo {year} {1961})}\BibitemShut {NoStop}%
		\bibitem [{\citenamefont {Ayaz}\ \emph {et~al.}(2022)\citenamefont {Ayaz},
			\citenamefont {Scalfi}, \citenamefont {Dalton},\ and\ \citenamefont
			{Netz}}]{cihan2022_hybrid_gle}%
		\BibitemOpen
		\bibfield  {author} {\bibinfo {author} {\bibfnamefont {C.}~\bibnamefont
				{Ayaz}}, \bibinfo {author} {\bibfnamefont {L.}~\bibnamefont {Scalfi}},
			\bibinfo {author} {\bibfnamefont {B.~A.}\ \bibnamefont {Dalton}},\ and\
			\bibinfo {author} {\bibfnamefont {R.~R.}\ \bibnamefont {Netz}},\ }\bibfield
		{title} {\bibinfo {title} {Generalized langevin equation with a nonlinear
				potential of mean force and nonlinear memory friction from a hybrid
				projection scheme},\ }\href {https://doi.org/10.1103/PhysRevE.105.054138}
		{\bibfield  {journal} {\bibinfo  {journal} {Phys. Rev. E}\ }\textbf {\bibinfo
				{volume} {105}},\ \bibinfo {pages} {054138} (\bibinfo {year}
			{2022})}\BibitemShut {NoStop}%
		\bibitem [{\citenamefont {Mizuno}\ \emph {et~al.}(2007)\citenamefont {Mizuno},
			\citenamefont {Tardin}, \citenamefont {Schmidt},\ and\ \citenamefont
			{MacKintosh}}]{mizuno2007nonequilibrium}%
		\BibitemOpen
		\bibfield  {author} {\bibinfo {author} {\bibfnamefont {D.}~\bibnamefont
				{Mizuno}}, \bibinfo {author} {\bibfnamefont {C.}~\bibnamefont {Tardin}},
			\bibinfo {author} {\bibfnamefont {C.~F.}\ \bibnamefont {Schmidt}},\ and\
			\bibinfo {author} {\bibfnamefont {F.~C.}\ \bibnamefont {MacKintosh}},\
		}\bibfield  {title} {\bibinfo {title} {Nonequilibrium mechanics of active
				cytoskeletal networks},\ }\href@noop {} {\bibfield  {journal} {\bibinfo
				{journal} {Science}\ }\textbf {\bibinfo {volume} {315}},\ \bibinfo {pages}
			{370} (\bibinfo {year} {2007})}\BibitemShut {NoStop}%
		\bibitem [{\citenamefont {Netz}(2018)}]{netz2018non_equilibrium}%
		\BibitemOpen
		\bibfield  {author} {\bibinfo {author} {\bibfnamefont {R.~R.}\ \bibnamefont
				{Netz}},\ }\bibfield  {title} {\bibinfo {title} {Fluctuation-dissipation
				relation and stationary distribution of an exactly solvable many-particle
				model for active biomatter far from equilibrium},\ }\href@noop {} {\bibfield
			{journal} {\bibinfo  {journal} {The Journal of chemical physics}\ }\textbf
			{\bibinfo {volume} {148}},\ \bibinfo {pages} {185101} (\bibinfo {year}
			{2018})}\BibitemShut {NoStop}%
		\bibitem [{\citenamefont {Netz}(2023)}]{roland_neq_2023b}%
		\BibitemOpen
		\bibfield  {author} {\bibinfo {author} {\bibfnamefont {R.~R.}\ \bibnamefont
				{Netz}},\ }\href {https://arxiv.org/abs/2310.08886} {\bibinfo {title}
			{Multi-point distribution for gaussian non-equilibrium non-markovian
				observables}} (\bibinfo {year} {2023})\BibitemShut {NoStop}%
		\bibitem [{\citenamefont {Selmeczi}\ \emph {et~al.}(2005)\citenamefont
			{Selmeczi}, \citenamefont {Mosler}, \citenamefont {Hagedorn}, \citenamefont
			{Larsen},\ and\ \citenamefont {Flyvbjerg}}]{selmeczi2005cell}%
		\BibitemOpen
		\bibfield  {author} {\bibinfo {author} {\bibfnamefont {D.}~\bibnamefont
				{Selmeczi}}, \bibinfo {author} {\bibfnamefont {S.}~\bibnamefont {Mosler}},
			\bibinfo {author} {\bibfnamefont {P.~H.}\ \bibnamefont {Hagedorn}}, \bibinfo
			{author} {\bibfnamefont {N.~B.}\ \bibnamefont {Larsen}},\ and\ \bibinfo
			{author} {\bibfnamefont {H.}~\bibnamefont {Flyvbjerg}},\ }\bibfield  {title}
		{\bibinfo {title} {Cell motility as persistent random motion: theories from
				experiments},\ }\href@noop {} {\bibfield  {journal} {\bibinfo  {journal}
				{Biophysical journal}\ }\textbf {\bibinfo {volume} {89}},\ \bibinfo {pages}
			{912} (\bibinfo {year} {2005})}\BibitemShut {NoStop}%
		\bibitem [{\citenamefont {Flyvbjerg}\ and\ \citenamefont
			{Petersen}(1989)}]{flyvbjerg1989error}%
		\BibitemOpen
		\bibfield  {author} {\bibinfo {author} {\bibfnamefont {H.}~\bibnamefont
				{Flyvbjerg}}\ and\ \bibinfo {author} {\bibfnamefont {H.~G.}\ \bibnamefont
				{Petersen}},\ }\bibfield  {title} {\bibinfo {title} {Error estimates on
				averages of correlated data},\ }\href@noop {} {\bibfield  {journal} {\bibinfo
				{journal} {The Journal of Chemical Physics}\ }\textbf {\bibinfo {volume}
				{91}},\ \bibinfo {pages} {461} (\bibinfo {year} {1989})}\BibitemShut
		{NoStop}%
		\bibitem [{\citenamefont {Snijder}\ and\ \citenamefont
			{Pelkmans}(2011)}]{snijder2011origins}%
		\BibitemOpen
		\bibfield  {author} {\bibinfo {author} {\bibfnamefont {B.}~\bibnamefont
				{Snijder}}\ and\ \bibinfo {author} {\bibfnamefont {L.}~\bibnamefont
				{Pelkmans}},\ }\bibfield  {title} {\bibinfo {title} {Origins of regulated
				cell-to-cell variability},\ }\href@noop {} {\bibfield  {journal} {\bibinfo
				{journal} {Nature reviews Molecular cell biology}\ }\textbf {\bibinfo
				{volume} {12}},\ \bibinfo {pages} {119} (\bibinfo {year} {2011})}\BibitemShut
		{NoStop}%
		\bibitem [{\citenamefont {Spencer}\ \emph {et~al.}(2009)\citenamefont
			{Spencer}, \citenamefont {Gaudet}, \citenamefont {Albeck}, \citenamefont
			{Burke},\ and\ \citenamefont {Sorger}}]{spencer2009non}%
		\BibitemOpen
		\bibfield  {author} {\bibinfo {author} {\bibfnamefont {S.~L.}\ \bibnamefont
				{Spencer}}, \bibinfo {author} {\bibfnamefont {S.}~\bibnamefont {Gaudet}},
			\bibinfo {author} {\bibfnamefont {J.~G.}\ \bibnamefont {Albeck}}, \bibinfo
			{author} {\bibfnamefont {J.~M.}\ \bibnamefont {Burke}},\ and\ \bibinfo
			{author} {\bibfnamefont {P.~K.}\ \bibnamefont {Sorger}},\ }\bibfield  {title}
		{\bibinfo {title} {Non-genetic origins of cell-to-cell variability in
				trail-induced apoptosis},\ }\href@noop {} {\bibfield  {journal} {\bibinfo
				{journal} {Nature}\ }\textbf {\bibinfo {volume} {459}},\ \bibinfo {pages}
			{428} (\bibinfo {year} {2009})}\BibitemShut {NoStop}%
		\bibitem [{\citenamefont {Netz}(2024{\natexlab{b}})}]{NetzFilter2024}%
		\BibitemOpen
		\bibfield  {author} {\bibinfo {author} {\bibfnamefont {R.~R.}\ \bibnamefont
				{Netz}},\ }\bibfield  {title} {\bibinfo {title} {Temporal coarse-graining and
				elimination of slow dynamics with the generalized langevin equation for
				time-filtered observables},\ }\href@noop {} {\bibfield  {journal} {\bibinfo
				{journal} {arXiv preprint arXiv:2409.12429}\ } (\bibinfo {year}
			{2024}{\natexlab{b}})}\BibitemShut {NoStop}%
		\bibitem [{\citenamefont {Bechinger}\ \emph {et~al.}(2016)\citenamefont
			{Bechinger}, \citenamefont {Di~Leonardo}, \citenamefont {L{\"o}wen},
			\citenamefont {Reichhardt}, \citenamefont {Volpe},\ and\ \citenamefont
			{Volpe}}]{bechinger2016active}%
		\BibitemOpen
		\bibfield  {author} {\bibinfo {author} {\bibfnamefont {C.}~\bibnamefont
				{Bechinger}}, \bibinfo {author} {\bibfnamefont {R.}~\bibnamefont
				{Di~Leonardo}}, \bibinfo {author} {\bibfnamefont {H.}~\bibnamefont
				{L{\"o}wen}}, \bibinfo {author} {\bibfnamefont {C.}~\bibnamefont
				{Reichhardt}}, \bibinfo {author} {\bibfnamefont {G.}~\bibnamefont {Volpe}},\
			and\ \bibinfo {author} {\bibfnamefont {G.}~\bibnamefont {Volpe}},\ }\bibfield
		{title} {\bibinfo {title} {Active particles in complex and crowded
				environments},\ }\href@noop {} {\bibfield  {journal} {\bibinfo  {journal}
				{Reviews of Modern Physics}\ }\textbf {\bibinfo {volume} {88}},\ \bibinfo
			{pages} {045006} (\bibinfo {year} {2016})}\BibitemShut {NoStop}%
		\bibitem [{\citenamefont {Dieterich}\ \emph {et~al.}(2008)\citenamefont
			{Dieterich}, \citenamefont {Klages}, \citenamefont {Preuss},\ and\
			\citenamefont {Schwab}}]{dieterich2008anomalous_klages}%
		\BibitemOpen
		\bibfield  {author} {\bibinfo {author} {\bibfnamefont {P.}~\bibnamefont
				{Dieterich}}, \bibinfo {author} {\bibfnamefont {R.}~\bibnamefont {Klages}},
			\bibinfo {author} {\bibfnamefont {R.}~\bibnamefont {Preuss}},\ and\ \bibinfo
			{author} {\bibfnamefont {A.}~\bibnamefont {Schwab}},\ }\bibfield  {title}
		{\bibinfo {title} {Anomalous dynamics of cell migration},\ }\href@noop {}
		{\bibfield  {journal} {\bibinfo  {journal} {Proceedings of the National
					Academy of Sciences}\ }\textbf {\bibinfo {volume} {105}},\ \bibinfo {pages}
			{459} (\bibinfo {year} {2008})}\BibitemShut {NoStop}%
		\bibitem [{\citenamefont {Sadhu}\ \emph {et~al.}(2023)\citenamefont {Sadhu},
			\citenamefont {Igli{\v{c}}},\ and\ \citenamefont {Gov}}]{sadhu2023minimal}%
		\BibitemOpen
		\bibfield  {author} {\bibinfo {author} {\bibfnamefont {R.~K.}\ \bibnamefont
				{Sadhu}}, \bibinfo {author} {\bibfnamefont {A.}~\bibnamefont {Igli{\v{c}}}},\
			and\ \bibinfo {author} {\bibfnamefont {N.~S.}\ \bibnamefont {Gov}},\
		}\bibfield  {title} {\bibinfo {title} {A minimal cell model for
				lamellipodia-based cellular dynamics and migration},\ }\href@noop {}
		{\bibfield  {journal} {\bibinfo  {journal} {Journal of Cell Science}\
			}\textbf {\bibinfo {volume} {136}},\ \bibinfo {pages} {jcs260744} (\bibinfo
			{year} {2023})}\BibitemShut {NoStop}%
		\bibitem [{\citenamefont {Heyn}\ \emph {et~al.}(2024)\citenamefont {Heyn},
			\citenamefont {Atienza~Juanatey}, \citenamefont {Falcke},\ and\ \citenamefont
			{Raedler}}]{heyn2024cell}%
		\BibitemOpen
		\bibfield  {author} {\bibinfo {author} {\bibfnamefont {J.}~\bibnamefont
				{Heyn}}, \bibinfo {author} {\bibfnamefont {M.}~\bibnamefont
				{Atienza~Juanatey}}, \bibinfo {author} {\bibfnamefont {M.}~\bibnamefont
				{Falcke}},\ and\ \bibinfo {author} {\bibfnamefont {J.}~\bibnamefont
				{Raedler}},\ }\bibfield  {title} {\bibinfo {title} {Cell-mechanical parameter
				estimation from 1d cell trajectories using simulation-based inference},\
		}\href@noop {} {\bibfield  {journal} {\bibinfo  {journal} {bioRxiv}\ ,\
				\bibinfo {pages} {2024}} (\bibinfo {year} {2024})}\BibitemShut {NoStop}%
		\bibitem [{\citenamefont {Pesarin}\ and\ \citenamefont
			{Salmaso}(2010)}]{pesarin2010permutation}%
		\BibitemOpen
		\bibfield  {author} {\bibinfo {author} {\bibfnamefont {F.}~\bibnamefont
				{Pesarin}}\ and\ \bibinfo {author} {\bibfnamefont {L.}~\bibnamefont
				{Salmaso}},\ }\bibfield  {title} {\bibinfo {title} {The permutation testing
				approach: a review},\ }\href@noop {} {\bibfield  {journal} {\bibinfo
				{journal} {Statistica}\ }\textbf {\bibinfo {volume} {70}},\ \bibinfo {pages}
			{481} (\bibinfo {year} {2010})}\BibitemShut {NoStop}%
		\bibitem [{\citenamefont {Wallert}\ \emph {et~al.}(2020)\citenamefont
			{Wallert}, \citenamefont {Nie}, \citenamefont {Anilkumar}, \citenamefont
			{Abbina}, \citenamefont {Bhatia}, \citenamefont {Ludwig}, \citenamefont
			{Kizhakkedathu}, \citenamefont {Haag},\ and\ \citenamefont
			{Block}}]{wallert2020mucin}%
		\BibitemOpen
		\bibfield  {author} {\bibinfo {author} {\bibfnamefont {M.}~\bibnamefont
				{Wallert}}, \bibinfo {author} {\bibfnamefont {C.}~\bibnamefont {Nie}},
			\bibinfo {author} {\bibfnamefont {P.}~\bibnamefont {Anilkumar}}, \bibinfo
			{author} {\bibfnamefont {S.}~\bibnamefont {Abbina}}, \bibinfo {author}
			{\bibfnamefont {S.}~\bibnamefont {Bhatia}}, \bibinfo {author} {\bibfnamefont
				{K.}~\bibnamefont {Ludwig}}, \bibinfo {author} {\bibfnamefont {J.~N.}\
				\bibnamefont {Kizhakkedathu}}, \bibinfo {author} {\bibfnamefont
				{R.}~\bibnamefont {Haag}},\ and\ \bibinfo {author} {\bibfnamefont
				{S.}~\bibnamefont {Block}},\ }\bibfield  {title} {\bibinfo {title}
			{Mucin-inspired, high molecular weight virus binding inhibitors show biphasic
				binding behavior to influenza a viruses},\ }\href@noop {} {\bibfield
			{journal} {\bibinfo  {journal} {Small}\ }\textbf {\bibinfo {volume} {16}},\
			\bibinfo {pages} {2004635} (\bibinfo {year} {2020})}\BibitemShut {NoStop}%
		\bibitem [{\citenamefont {M{\"u}ller}\ \emph {et~al.}(2019)\citenamefont
			{M{\"u}ller}, \citenamefont {Lauster}, \citenamefont {Wildenauer},
			\citenamefont {Herrmann},\ and\ \citenamefont {Block}}]{muller2019mobility}%
		\BibitemOpen
		\bibfield  {author} {\bibinfo {author} {\bibfnamefont {M.}~\bibnamefont
				{M{\"u}ller}}, \bibinfo {author} {\bibfnamefont {D.}~\bibnamefont {Lauster}},
			\bibinfo {author} {\bibfnamefont {H.~H.}\ \bibnamefont {Wildenauer}},
			\bibinfo {author} {\bibfnamefont {A.}~\bibnamefont {Herrmann}},\ and\
			\bibinfo {author} {\bibfnamefont {S.}~\bibnamefont {Block}},\ }\bibfield
		{title} {\bibinfo {title} {Mobility-based quantification of multivalent
				virus-receptor interactions: new insights into influenza a virus binding
				mode},\ }\href@noop {} {\bibfield  {journal} {\bibinfo  {journal} {Nano
					letters}\ }\textbf {\bibinfo {volume} {19}},\ \bibinfo {pages} {1875}
			(\bibinfo {year} {2019})}\BibitemShut {NoStop}%
		\bibitem [{\citenamefont {Kerkhoff}\ and\ \citenamefont
			{Block}(2020)}]{kerkhoff2020analysis}%
		\BibitemOpen
		\bibfield  {author} {\bibinfo {author} {\bibfnamefont {Y.}~\bibnamefont
				{Kerkhoff}}\ and\ \bibinfo {author} {\bibfnamefont {S.}~\bibnamefont
				{Block}},\ }\bibfield  {title} {\bibinfo {title} {Analysis and refinement of
				2d single-particle tracking experiments},\ }\href@noop {} {\bibfield
			{journal} {\bibinfo  {journal} {Biointerphases}\ }\textbf {\bibinfo {volume}
				{15}} (\bibinfo {year} {2020})}\BibitemShut {NoStop}%
		\bibitem [{\citenamefont {von Hansen}\ \emph {et~al.}(2012)\citenamefont {von
				Hansen}, \citenamefont {Mehlich}, \citenamefont {Pelz}, \citenamefont
			{Rief},\ and\ \citenamefont {Netz}}]{von2012auto}%
		\BibitemOpen
		\bibfield  {author} {\bibinfo {author} {\bibfnamefont {Y.}~\bibnamefont {von
					Hansen}}, \bibinfo {author} {\bibfnamefont {A.}~\bibnamefont {Mehlich}},
			\bibinfo {author} {\bibfnamefont {B.}~\bibnamefont {Pelz}}, \bibinfo {author}
			{\bibfnamefont {M.}~\bibnamefont {Rief}},\ and\ \bibinfo {author}
			{\bibfnamefont {R.~R.}\ \bibnamefont {Netz}},\ }\bibfield  {title} {\bibinfo
			{title} {Auto-and cross-power spectral analysis of dual trap optical tweezer
				experiments using bayesian inference},\ }\href@noop {} {\bibfield  {journal}
			{\bibinfo  {journal} {Review of Scientific Instruments}\ }\textbf {\bibinfo
				{volume} {83}} (\bibinfo {year} {2012})}\BibitemShut {NoStop}%
		\bibitem [{\citenamefont {Schreiber}\ \emph {et~al.}(2016)\citenamefont
			{Schreiber}, \citenamefont {Segerer}, \citenamefont {Wagner}, \citenamefont
			{Roidl},\ and\ \citenamefont {R{\"a}dler}}]{schreiber2016ring}%
		\BibitemOpen
		\bibfield  {author} {\bibinfo {author} {\bibfnamefont {C.}~\bibnamefont
				{Schreiber}}, \bibinfo {author} {\bibfnamefont {F.~J.}\ \bibnamefont
				{Segerer}}, \bibinfo {author} {\bibfnamefont {E.}~\bibnamefont {Wagner}},
			\bibinfo {author} {\bibfnamefont {A.}~\bibnamefont {Roidl}},\ and\ \bibinfo
			{author} {\bibfnamefont {J.~O.}\ \bibnamefont {R{\"a}dler}},\ }\bibfield
		{title} {\bibinfo {title} {Ring-shaped microlanes and chemical barriers as a
				platform for probing single-cell migration},\ }\href@noop {} {\bibfield
			{journal} {\bibinfo  {journal} {Scientific reports}\ }\textbf {\bibinfo
				{volume} {6}},\ \bibinfo {pages} {26858} (\bibinfo {year}
			{2016})}\BibitemShut {NoStop}%
		\bibitem [{\citenamefont {Blair}\ and\ \citenamefont
			{Dufresne}(2008)}]{Matlabtrack}%
		\BibitemOpen
		\bibfield  {author} {\bibinfo {author} {\bibfnamefont {D.}~\bibnamefont
				{Blair}}\ and\ \bibinfo {author} {\bibfnamefont {E.}~\bibnamefont
				{Dufresne}},\ }\bibfield  {title} {\bibinfo {title} {The matlab particle
				tracking code repository},\ }\href
		{http://site.physics.georgetown.edu/matlab/code.html} {\bibfield  {journal}
			{\bibinfo  {journal} {Particle-tracking code available at http://physics.
					georgetown. edu/matlab}\ } (\bibinfo {year} {2008})}\BibitemShut {NoStop}%
	\end{thebibliography}

	%

\makeatletter\@input{aux_file.tex}\makeatother
\end{document}


\title{Supporting Information:\\Intrinsic cell-to-cell variance from experimental single-cell motility data}
	
	
	\author{Anton Klimek}
	\affiliation{Fachbereich Physik, Freie Universit{\"a}t Berlin, 14195 Berlin, Germany}

	\author{Johannes C. J. Heyn}
	\affiliation{Physics Faculty and Center for NanoScience, Ludwig Maximilians Universit{\"a}t, 80539 M{\"u}nchen, Germany}
	
	\author{Debasmita Mondal}
	\affiliation{Department of Physics, Indian Institute of Science, 560012 Bangalore, India}
	\affiliation{James Franck Intitute, University of Chicago, 60637 Chicago, USA}
	
	\author{Sophia Schwartz}
	\affiliation{Fachbereich Chemie und Biochemie, Freie Universit{\"a}t Berlin, 14195 Berlin, Germany}
	
	\author{Joachim O. Rädler}
	\affiliation{Physics Faculty and Center for NanoScience, Ludwig Maximilians Universit{\"a}t, 80539 M{\"u}nchen, Germany}
	
	\author{Prerna Sharma}
	\affiliation{Department of Physics, Indian Institute of Science, 560012 Bangalore, India} 
	\affiliation{Department of Bioengineering, Indian Institute of Science, 560012 Bangalore, India}
	
	\author{Stephan Block}
	\affiliation{Fachbereich Chemie und Biochemie, Freie Universit{\"a}t Berlin, 14195 Berlin, Germany}
	
	\author{Roland R. Netz}
	\affiliation{Fachbereich Physik, Freie Universit{\"a}t Berlin, 14195 Berlin, Germany}
	\email{rnetz@physik.fu-berlin.de}

	\maketitle
	\tableofcontents

	\section{Averaging velocities over populations can result in non-Gaussian distributions}
\begin{figure*}
		\centering
		\includegraphics[width=\textwidth]{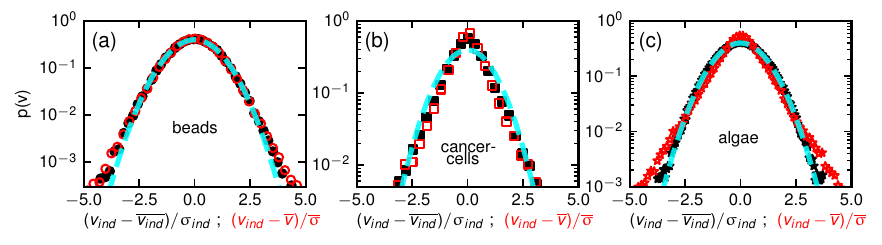}
		\caption{Average velocity distribution of all moving objects for the beads (a), cancer cells (b), algae (c).
		The black symbols correspond to the averages shown in Fig. \ref{fig_1}b,e,h and result from rescaling with respect to individual trajectory means $\overline{v_{\rm{ind}}}$ and standard deviations $\sigma_{\rm{ind}}$ according to the black x-axis label.
		Red symbols result from rescaling by the population mean $\overline{v}$ and standard deviation $\overline{\sigma}$ according to eq. \eqref{eq_v_rescale_population}.
		The dashed cyan line represents the standard normal distribution.}
		\label{fig_hist_nongauss}
	\end{figure*}

	We show in the main text that the velocity distributions of the considered moving objects are Gaussian on the single-individual level in Fig. \ref{fig_1}.
	Even though all moving objects exhibit Gaussian distributions on the individual level, averaging over the entire population can lead to non-Gaussian effects.
	In Fig. \ref{fig_hist_nongauss} we show that averaging with respect to the population mean velocity $\overline{v}$ and the population standard deviation $\overline{\sigma}$ according to
	\begin{equation}
		\label{eq_v_rescale_population}
		v = (v_{\rm{ind}} - \overline{v})/\overline{\sigma}
	\end{equation}
	can lead to deviations from Gaussian behavior.
	These deviations are negligible for the beads shown in Fig. \ref{fig_hist_nongauss}a.
	For the cancer cells deviations from Gaussian behavior are slightly increased when using eq. \eqref{eq_v_rescale_population} compared to the rescaling on the individual level, as shown in Fig. \ref{fig_hist_nongauss}b.
	For the algae cells the deviations from Gaussian behavior are quite drastic, when rescaling according to eq. \eqref{eq_v_rescale_population}.
	The results of Fig. \ref{fig_hist_nongauss} showcase that it is important to analyze motility data on the single-individual level and that averaging of quantities over individual moving objects can lead to non-Gaussian effects.
	
	\section{Memory kernel extraction}
	\label{sec_kernel_extraction}
	Multiplying the GLE eq. \eqref{eq_gle} by the initial velocity $\dot{x}(t_0)$, averaging over the random force
	and integrating from $t_0$ to $t$ leads to
	\begin{equation}
		\label{eq_volterra}
		(C_{vv}(t)-C_{vv}(0)) = -\int_{0}^{t} C_{vv}(s) G(t-s) ds \,,
	\end{equation}
	where we used that  $\langle \dot{x}(t_0) F_R(t)\rangle =0 $ \cite{mori_transport_1965,zwanzig1961memory,cihan2022_hybrid_gle},
	set  $t_0=0$  and introduced the integral kernel 
	\begin{equation}
		\label{eq_G_def}
		G(t) = \int_{0}^{t} \Gamma (s) ds \,.
	\end{equation}
	In order to invert eq. \eqref{eq_volterra}, we discretize it.
	Since $C_{vv}(t)$ is even while $G(t)$  is odd, we
	discretize  $G(t)$ on half steps and $C_{vv}(t)$ on full steps and obtain \cite{mitterwallner_non-markovian_2020} 
	\begin{equation}
		\label{eq_G_half_step}
		G_{i+1/2} = \frac{2(C_{vv}^0 -C_{vv}^{i+1})}{\Delta(C_{vv}^1 + C_{vv}^0)} - \sum_{j=1}^{i} G_{i-j+1/2}\frac{C_{vv}^{j+1}+C_{vv}^j}{C_{vv}^1+C_{vv}^0} \,.
	\end{equation}
	The kernel $\Gamma_i$ is  obtained by  the discrete derivative $\Gamma_i=\frac{G_{i+1/2} + G_{i-1/2}}{\Delta}$
	with the initial value  $\Gamma_0 = 2G_{1/2}/\Delta$.
	
	\section{Goodness of fit to individual trajectory data}
	\label{sec_goodness_of_fit}
\begin{figure*}
		\centering
		\includegraphics[width=\textwidth]{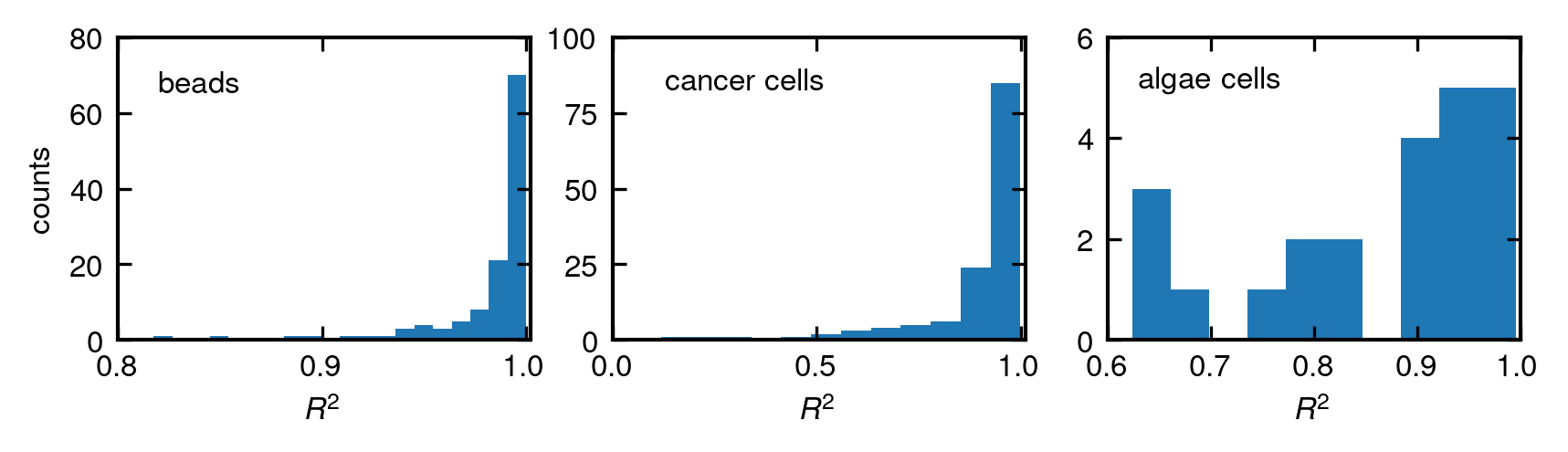}
		
		\caption{Histogram of the $R^2$ values defined in eq. \eqref{eq_def_rsq} for the fits of the respective model eqs. \eqref{eq_gle_diffusion_lim},\eqref{eq_kern_prw},\eqref{eq_kern_osc}  to individual trajectories for polystyrene beads (a), breast cancer cells (b), algae cells (c).}
		\label{fig_SI_hist_rsq}
	\end{figure*}

	In Fig. \ref{fig_3} of the main text we show single-trajectory fit results of the models eqs. \eqref{eq_kern_osc}, \eqref{eq_kern_prw}, \eqref{eq_gle_diffusion_lim} to randomly chosen single trajectory data.
	Since we cannot show the fit results of all individuals, we instead show the distribution of $R^2$ values for all fits in Fig. \ref{fig_SI_hist_rsq}.
	Here, $R^2$ is defined using the cost function $E_{\rm{cost}}$ of eq. \eqref{eq_cost_function} as
	\begin{equation}
		\label{eq_def_rsq}
		R^2 = 1 - \frac{E_{\rm{cost}}}{\sum_{i=0}^{n-1} (C_{vv}^{\rm{exp}}(i \Delta)-\overline{C_{vv}^{\rm{exp}}})^2}\,,
	\end{equation}
	where $\overline{C_{vv}^{\rm{exp}}}$ denotes the mean value of the VACF over the $n$ fitted time steps.
	Therefore, $R^2$ gives an estimate for the goodness of the fit, where a value of zero corresponds to a bad fit that is equivalent to assuming the function to be the mean everywhere and a value of one corresponds to perfect agreement between the fit and the data.
	
\begin{figure*}
		\centering
		\includegraphics[width=\textwidth]{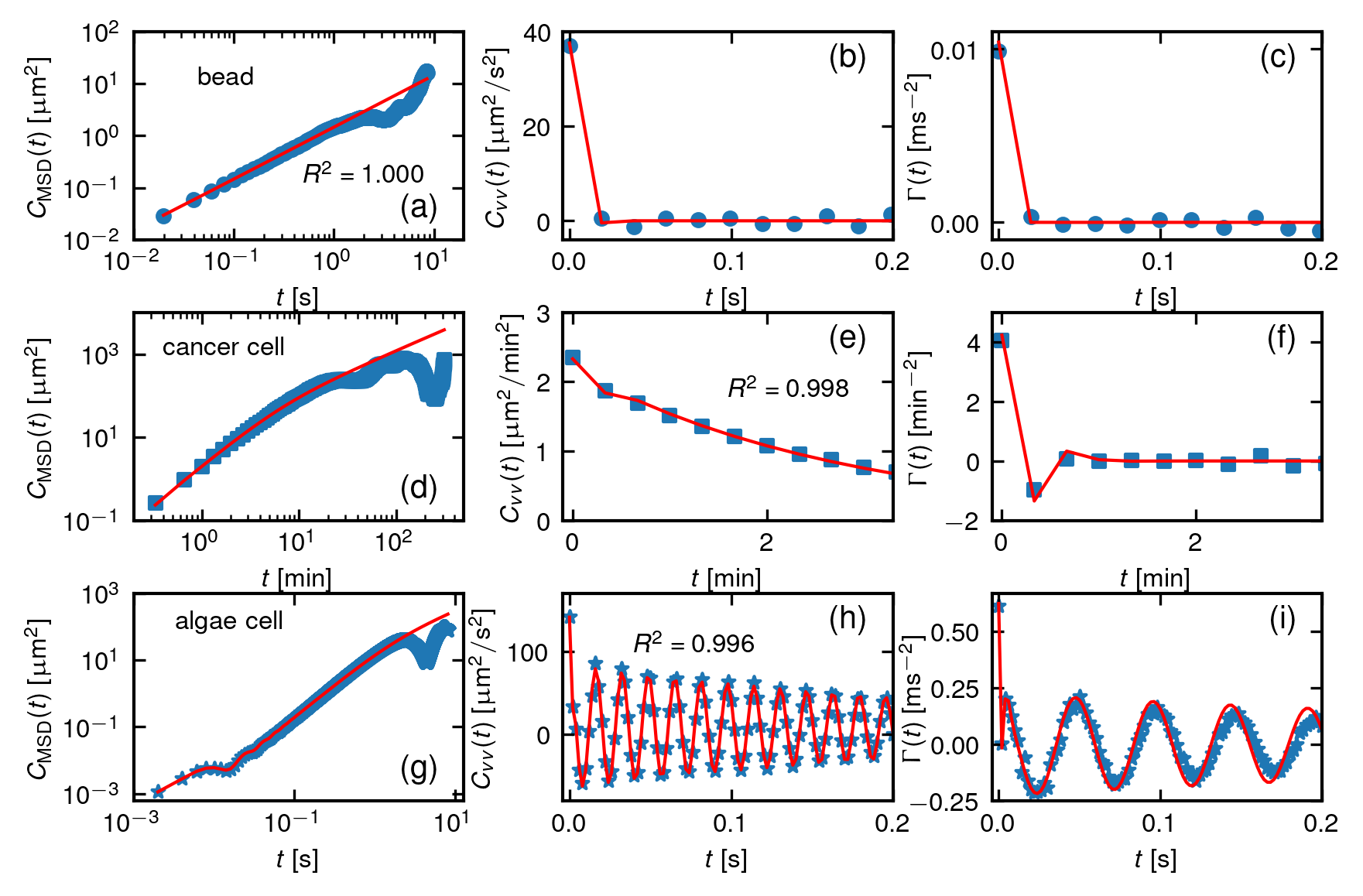}
		
		\caption{Experimental results for the  MSD, $C_{\textrm{MSD}}(t)$, VACF, $C_{vv}(t)$, and friction kernel $\Gamma(t)$,  of a single bead {(a)-(c)},  a single cancer cell {(d)-(f)} and a single algae cell {(g)-(i)} (blue symbols), chosen as the objects with the highest $R^2$ value, which is given in the respective plot.
		The red lines denote the fit result.}
		\label{fig_best_fits}
	\end{figure*}

	Naturally, for noisy data obtained from experiments, there is some spread in the $R^2$ values, as shown in Fig. \ref{fig_SI_hist_rsq}.
	Nevertheless, it can be seen that most of the $R^2$ values of our extraction are close to one, thus indicating that our different fit models describe the respective data accurately on the individual trajectory level.
	To further undermine this point, we show the fit results with the highest $R^2$ value for the respective individuals in Fig. \ref{fig_best_fits}.
	Here, the data agrees perfectly with the fit results.
	The $R^2$ values of the respective fits for the randomly chosen individuals in Figs. \ref{fig_3}a,e,h are $1.000$, $0.964$, $0.953$.
	
	\section{Effective kernel follows uniquely from correlation function}\label{sec_mapping_SI}
	Fourier transformation of eq. \eqref{eq_gle} and eq. \eqref{eq_fdt_neq} leads to
	\begin{equation}
		\label{eq_ft_gle_neq}
		\tilde{v}(\omega)=\frac{\tilde{F}_R(\omega)}{\tilde{\Gamma}^+_v(\omega) +i\omega}
	\end{equation}
	with the single-sided Fourier transform defined as 
	$\tilde{\Gamma}^+_v(\omega) = \int_{0}^{\infty} e^{-i\omega t} {\Gamma}_v(t) dt$ and 
	\begin{equation}
		\label{eq_ft_force_corr}
		\langle  \tilde{F}_R(\omega) \tilde{F}_R(\omega') \rangle = 2\pi \delta(\omega+\omega')\tilde{\Gamma}_R(\omega')\,.
	\end{equation}
	From the Fourier transform of the VACF
	\begin{equation}
		\begin{split}
			\label{eq_ft_vacf}
			&\tilde{C}_{vv}(\omega) = \int_{-\infty}^{\infty} dt e^{-i\omega t}\langle v(0)v(t)  \rangle \\
			&= \int_{-\infty}^{\infty}e^{-i\omega t} dt \int_{\infty}^{\infty}e^{i\omega t} \frac{d\omega}{2\pi} \int_{-\infty}^{\infty} \frac{d\omega}{2\pi} 
			\langle \tilde{v}(\omega) \tilde{v}(\omega')  \rangle
		\end{split}
	\end{equation}
	we obtain by inserting eqs. \eqref{eq_ft_gle_neq} and \eqref{eq_ft_force_corr} 
	\begin{equation}
		\label{eq_ft_vacf_final}
		\tilde{C}_{vv}(\omega) = \frac{\tilde{\Gamma}_R(\omega)}{(\tilde{\Gamma}_v^+(\omega) +i\omega) (\tilde{\Gamma}_v^+(-\omega) -i\omega)} \,.
	\end{equation}
	Equating the non-equilibrium and the surrogate VACF with $\Gamma_R(t)/B=\Gamma_v(\mid t \mid)=\Gamma(\mid t \mid)$ leads to 
	\begin{equation}
		\label{eq_condition_eq_neq2}
		\frac{\tilde{\Gamma}_R(\omega)}{\mid\tilde{\Gamma}_v^+(\omega) +i\omega\mid^2} =
		\frac{B \tilde{\Gamma}(\omega)}{\mid\tilde{\Gamma}^+(\omega) +i\omega\mid^2} \,.
	\end{equation}
	
	Multiplying eq. \eqref{eq_gle}
	by $v(t_0)$, inserting $\Gamma(\vert t\vert)=\Gamma_R(t)/B=\Gamma_v(\vert t \vert)$ and averaging the entire equation leads to the Volterra equation
	\begin{equation}
		\label{eq_volterra_pot}
		\dfrac{d}{dt}C_{vv}(t) = -\int_{0}^{t}\Gamma(t-s)C_{vv}(s)ds \,,
	\end{equation}
	where we used that the random force is not correlated with the initial velocity at the projection time, i.e. $\langle v(t_0)F_R(t) \rangle = 0$, and set $t_0=0$.
	For a given $\Gamma(t)$ one can find a solution of eq. \eqref{eq_volterra_pot} in terms of $C_{vv}(t)$.
	Inversely, for given correlation function $C_{vv}(t)$ one can solve eq. \eqref{eq_volterra_pot} in terms of the friction kernel $\Gamma(t)$ by Laplace transformation.
	With the definition of the Laplace transform of a function $f(t)$ 
	\begin{equation}
		\hat{f}(q) = \int_0^{\infty}f(t) e^{-qt}dt,
		\label{eq_def_laplace_transform}
	\end{equation}
	the unique solution for $\Gamma(t)$ in Laplace space is given by
	\begin{equation}
		\label{eq_solution_volterra_gamma_laplace}
		\hat{\Gamma}(q)= \left( \frac{C_{vv}(0)}{\hat{C}_{vv}(q)} - q \right) \,.
	\end{equation}
	The existence of a unique friction kernel for any given input correlation function $C_{vv}(t)$
	assures that one can always find an effective friction kernel $\Gamma(t)$ that, when employed in the GLE and using  $\Gamma(\vert t\vert) = \Gamma_R(t)/B = \Gamma_v(\vert t\vert)$,
	reproduces the two-point correlation functions.
	Thus, every non-equilibrium model described by  eq. \eqref{eq_gle} 
	with  $ \Gamma_R(t)/B \neq \Gamma_v(\vert t\vert)$
	can be mapped on an effective model with $\Gamma(\vert t\vert) = \Gamma_R(t)/B = \Gamma_v(\vert t\vert)$ determined by eq. \eqref{eq_solution_volterra_gamma_laplace}, because the Green's function, 
	which completely  describes a Gaussian process,
	is solely given in terms of the positional two-point correlation function \cite{roland_neq_2023b,klimek_data-driven_2024}.

	\section{Bead radius variance determined by atomic force microscopy}
\begin{figure*}
		\centering
		\includegraphics[width=0.5\textwidth]{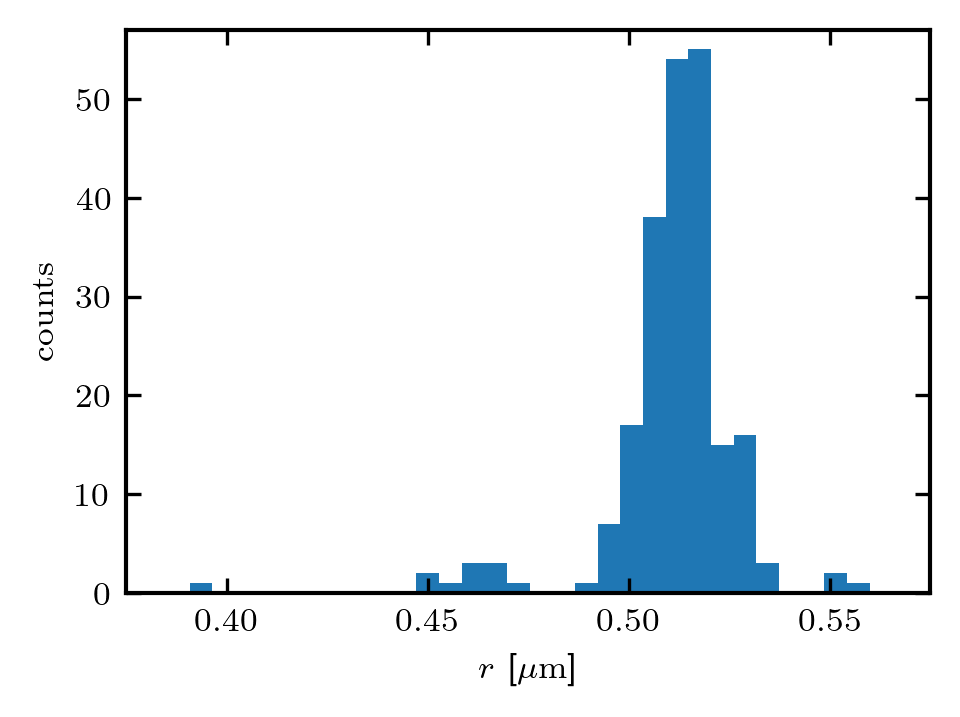}
		
		\caption{Distribution of measured bead radii via AFM, where the exact sample preparation and AFM settings are explained in the text.}
		\label{fig_radius_afm}
	\end{figure*}

	In order to estimate the bead radius variance independently of the motion in water, we determine the radius variance from $220$ beads via atomic force microscopy (AFM).
	The radius distribution is shown in Fig. \ref{fig_radius_afm}, where the mean radius is $\langle r \rangle=0.51\unit{\mu m}$ and the standard deviation is $\Delta r = 0.02\unit{\mu m}$.
	This is in line with the values of $\langle r \rangle=0.50\unit{\mu m}$ and standard deviation $\Delta r = 25\unit{nm}$ provided by the manufacturer.
	These values agree with the estimated radius variance from the motility data in the main text of $\Delta r = 86\pm63\unit{nm}$.
	
	The sample of beads is prepared on freshly cleaved mica. Before the AFM measurement the beads (FluoSpheres NeutrAvidin-Labeled Microspheres, $1.0\unit{\mu m}$ ThermoFisher Scientific) are diluted 1:100 in deionized water and sonicated for $10\unit{min}$ to
	reduce bead aggregation.
	The sample was then left to dry overnight.
	AFM images are obtained using a MultiMode 8 atomic force microscope
	(Veeco, Santa Barbara, CA) with a Nanoscope V controller and NanoScope
	software.
	Imaging is performed with TappingMode in air using AC160TS
	cantilevers (Asylum Research by Oxford Instruments, LOT$\#$753014).
	The AFM images with a frame size of $10\unit{\mu m} \times 10\unit{\mu m}$
	are recorded at a scan rate of $1\unit{Hz}$ over 256 lines, with an integral gain of 5 and
	proportional gain of 10, without z-limit.

	\section{Least-square fit parameter distributions of noisy exponentials: The median is more accurate than the mean}
\begin{figure*}
		\centering
		\includegraphics[width=\textwidth]{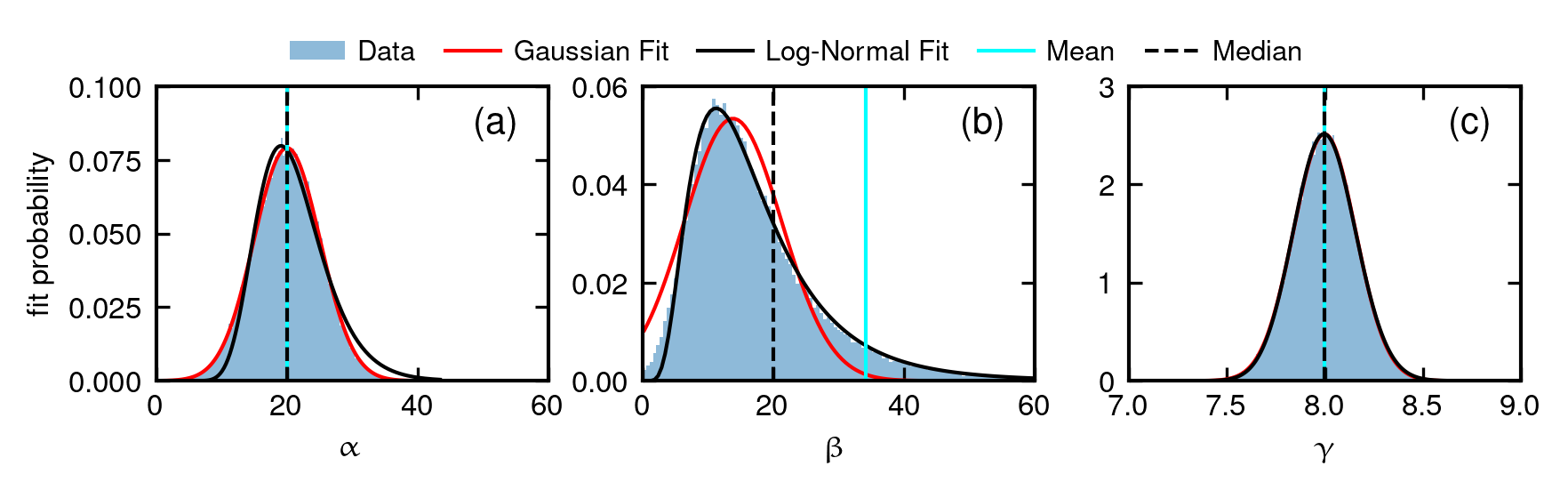}
		
		\caption{Distributions of fitted parameters (a) $\alpha$, (b) $\beta$, (c) $\gamma$ of $10^5$ fits of eq. \eqref{eq_fit_function_fit} to data sets of eq. \eqref{eq_fit_function} with input parameters $\alpha_0=20$, $\beta_0=20$, $\gamma_0=8$, $\langle \zeta(0) \zeta(t) = 5\delta(t)$ for $t$ between $0-100$ in steps of $0.1$. The mean and median of the data are presented as vertical lines and are given by $\langle \alpha\rangle=19.98$, $\alpha^{\rm{med}}=20.00$, $\langle\beta\rangle=34.18$, $\beta^{\rm{med}}=19.98$, $\langle\gamma\rangle=8.00$, $\gamma^{\rm{med}}=8.00$.}
		\label{fig_median_reasoning}
	\end{figure*}

	In Fig. \ref{fig_median_reasoning} we show the fit results of $10^5$ least-square fits to the noisy function
	\begin{equation}
		\label{eq_fit_function}
		f(t) = \alpha_0 e^{-\beta_0 t} + \gamma_0 + \zeta(t)
	\end{equation}
	with the fit function
	\begin{equation}
		\label{eq_fit_function_fit}
		f^{\rm{fit}}(t) = \alpha e^{-\beta t} + \gamma\,,
	\end{equation}
	where $\zeta(t)$ describes Gaussian noise with zero mean $\langle \zeta(t) \rangle = 0$ and variance $\langle \zeta(0) \zeta(t) = 5\delta(t)$.
	The form of eq. \eqref{eq_fit_function} is motivated by the fact that VACFs often contain exponentially decaying components.
	Since our model invokes fitting to VACFs with exponential components with a least-square fit according to eq. \eqref{eq_cost_function} in the main text, here, we discuss the resulting parameter distributions of such fits.
	The distribution of fitted values for the exponential prefactor $\alpha$ and the constant $\gamma$ are well described by Gaussian distributions, see Figs. \ref{fig_median_reasoning}a,c, whereas the distribution of the exponential factor $\beta$ is better described by a log-normal distribution, as shown in Fig. \ref{fig_median_reasoning}b, which has the form
	\begin{equation}
		p(\beta) = \frac{\exp(-\frac{(\ln(\beta) - \ln(\beta^{\rm{med}}))^2} {2\sigma^2})}  {\beta \sigma \sqrt{2\pi}}\,,
	\end{equation}
	where $\beta^{\rm{med}}$ denotes the median of the distribution and $\sigma$ determines the width of the distribution.
	
	It turns out that the actual input parameter of $\beta_0=20$ is better described by the median of the distribution of fitted parameters compared to the mean, which we show in Fig. \ref{fig_median_reasoning}b.
	For the parameters $\alpha$ and $\gamma$ the median and mean describe the actual input parameters equally well, as seen in Fig. \ref{fig_median_reasoning}a,c.
	As the median  accurately describes the true value of Gaussian distributed parameters as well as Log-normal distributed parameters from least-squares fits, we choose it as a representation of the center of distributions rather than the mean.
	
	\section{Analytical estimation of parameter variance}
	\label{sec_error_estimation}
	We compare our parameter variance estimate from simulations with analytical estimates.
	As an example, we estimate the mean squared velocity $B=\langle v^2 \rangle = C_{vv}(0)$ using the PRW model eq. \eqref{eq_kern_prw}.
	Due to our extraction of the model parameters from the VACF eq. \eqref{eq_vacf_fit}, the statistical error of the parameters relates closely to the statistical error of the VACF.
	We consider a trajectory composed of $n+1=L/\Delta$ positions which produces a VACF with $n$ values, where the mean squared velocity of the continuous trajectory is estimated by the finite difference velocities eq. \eqref{eq_central_diff_vel} as
	\begin{equation}
		\label{eq_def_Bestimator}	B_{\rm{est}}=\frac{1}{n}\sum_{i=1}^{n}v_i^2 \,.
	\end{equation}
	Here, it should be noted that this estimate deviates from the estimation of our fit model including localization noise effects eq. \eqref{eq_vacf_fit}.
	Especially in cases with localization noise, the estimate $B_{\rm{est}}$ in eq. \eqref{eq_def_Bestimator} overestimates the true value of $B$ as displacements are on average increased by the noise.
	On the other hand, a finite time step discretization leads to decreased values $B_{\rm{est}}$, as a finite time step does not capture instantaneous velocities but rather an average over the interval of the time step.
	Our extraction of parameters by a least squares fit of eq. \eqref{eq_vacf_fit} accounts for discretization and localization noise effects, but is not analytically tractable.
	
	The moments of $v$ follow from the velocity distribution $\rho(v)$ as 
	\begin{equation}
		\langle v^k \rangle = \int_{-\infty}^{\infty} v^k\rho(v) dv \,.
	\end{equation}
	We want to determine the dependence of the variance of our estimation $B_{\rm{est}}$ on the length of the trajectories
	\begin{equation}
		\label{eq_var_Best}
		\sigma^2(B_{\rm{est}})=\langle B_{\rm{est}}^2 \rangle -\langle B_{\rm{est}} \rangle^2\,.
	\end{equation}
	Inserting eq. \ref{eq_def_Bestimator} into eq. \ref{eq_var_Best} leads to
	\begin{eqnarray}
		\label{var_BN}
		&\sigma^2(B_{\rm{est}})= \nonumber \\
		&\frac{1}{n^2}\langle \sum_{i=1}^{n}v_i^2 \sum_{j}^{n}v_j^2 \rangle -(\frac{1}{n}\sum_{i=1}^{n} \langle v_i^2\rangle)^2 \\
		=&\frac{1}{n^2}(n\langle v^4\rangle +\sum_{i\neq j}^{n}\langle v_i^2 v_j^2\rangle  -n^2\langle v^2\rangle^2 ) \nonumber \,.
	\end{eqnarray}
	If the velocities at time $i$ and $j$ are independent of each other,
	the correlation of the squared velocities can be written as
		\begin{equation}
				\langle v_i^2 v_j^2\rangle=\langle v_i^2 \rangle \langle v_j^2 \rangle\, = \langle v^2\rangle^2,
			\end{equation}
	thus, the variance of $B_{\rm{est}}$ follows as
	\begin{equation}
		\label{varB_uncorr}
		\sigma^2(B_{\rm{est}}) = \frac{1}{n^2}\left(n\langle v^4\rangle + n(n-1)\langle v^2\rangle^2 -n^2\langle v^2\rangle^2\ \right) = \frac{1}{n}\left( \langle v^4 \rangle\ -\langle v^2 \rangle^2 \right)\,.
	\end{equation}
	In fact, the assumption of uncorrelated squared velocities is not true for most experimental trajectories, indicated by the non-zero values of the VACF for $t>0$, meaning one needs to consider the correlation between the squared velocities over time.
	This results in a different estimate for the variance of $B_{\rm{est}}$ \cite{flyvbjerg1989error}, which can be written as 
	\begin{align}
		\sigma^2(B_{\rm{est}}) &= \frac{1}{n}(\langle v^4 \rangle -\langle v^2\rangle^2 \\
		&+ 2\sum_{k=1}^{n-1}(1-k/n)\langle v^2_0 v^2_k\rangle) \nonumber \,,
	\end{align}
	where we used that the correlation function only depends on time differences $k=|i-j|$ and that the $k$th term $\langle v^2_0 v^2_k\rangle$ occurs $n-k$ times in the sum.
	If one does not know the analytic form of $C_{v^2v^2}(t) = \langle v^2(0)v^2(t) \rangle$, one can use the function obtained from the data similarly to eq. \ref{eq_vacf_from_data} for $C_{vv}(t)$, where one usually introduces a cutoff length $n_{\rm{cut}}$ to estimate the variance as
	\begin{equation}
		\label{eq_variance_estim_corr}
		\sigma^2(B_{\rm{est}})=\frac{1}{n} \left( C_{v^2v^2}^0 - (C_{vv}^0)^2 + 2\sum_{k=1}^{n_{\rm{cut}}}C_{v^2v^2}^k \right)
	\end{equation}
	in order to avoid using the estimated correlation function for large times $t>n_{\rm{cut}}\Delta$, where the noise is stronger due to less averaging, see eq. \ref{eq_vacf_from_data}.
	Similarly to the VACF eq. \eqref{eq_vacf_prw} one can calculate the correlation of the squared velocities for the PRW (the GLE eq. \eqref{eq_gle} with the instantaneous friction kernel eq. \eqref{eq_kern_prw}) starting from the formal solution for the velocity
	\begin{equation}
		v(t)=v(0)e^{-t/\tau_m} + \int_0^t F_R(s) e^{-(t-s)/\tau_m} ds\,.
	\end{equation}
	The squared velocity correlation function then takes the form
	\begin{equation}
		\langle v^2(0) v^2(t) \rangle = \left\langle v^2(0) \left[ v^2(0) e^{-2t/\tau_m} + 2 v(0) e^{-t/\tau_m} \int_0^t F_R(s) e^{-(t-s)/\tau_m} \, ds + \left( \int_0^t F_R(s) e^{-(t-s)/\tau_m} \, ds \right)^2 \right] \right\rangle\,.
	\end{equation}
	Inserting the first and second moment of the random force $\langle F_R(s) \rangle = 0$ and $\langle F_R(s) F_R(s') \rangle = 2B\delta(s-s')/\tau_m$ respectively and evaluating the integrals, one finds
	\begin{equation}
		\langle v^2(0) v^2(t) \rangle = 3 \langle v^2(0) \rangle^2 e^{-2t/\tau_m} + B \langle v^2(0) \rangle \left( 1 - e^{-2t/\tau_m} \right)\,,
	\end{equation}
	where we used that the fourth moment of a Gaussian is $\langle v(0)^4 \rangle = 3 \langle v^2(0) \rangle^2$.
	Now inserting $\langle v^2(0) \rangle = B$ yields
	\begin{equation}
		\label{eq_prw_vsqacf_analytic}
		C_{v^2v^2}(t) = B^2 (1+2e^{-2t/\tau_m}) \,.
	\end{equation}
	
	\begin{figure*}
		\centering
		\includegraphics[width=\textwidth]{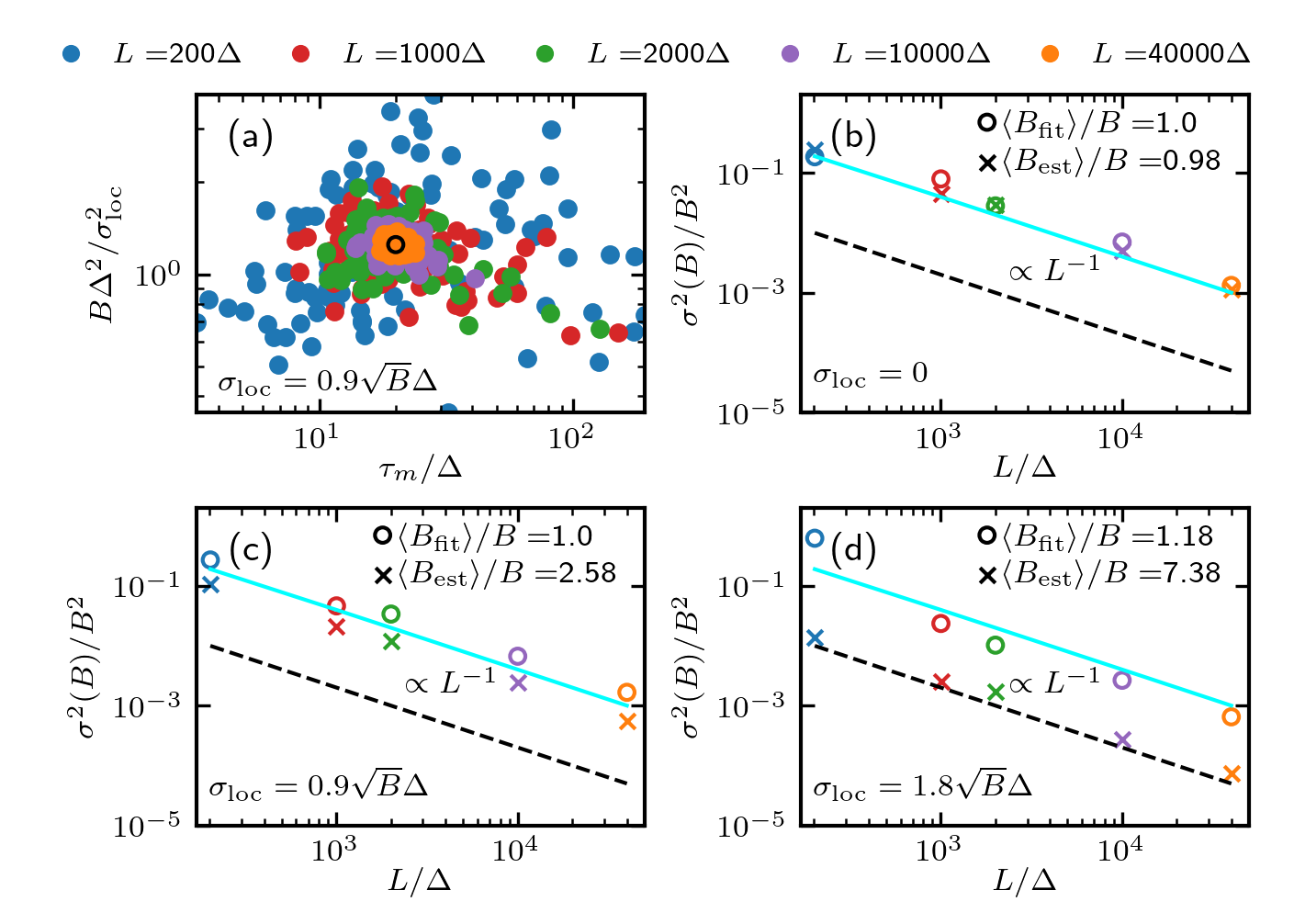}
		\caption{(a) Distributions of PRW parameters $B$, $\tau_m$ extracted from simulated trajectories of different lengths in different colors, where each distribution contains $N=100$ points and the input is shown as a black empty circle, $B\Delta^2/\sigma_{\rm{loc}} = 1.25$, $\tau_m/\Delta = 20$.
		(b)-(d) The variance of the estimator for $B$ in eq. \eqref{eq_def_Bestimator} is shown as crosses, the estimation of $B$ by our fit eq. \eqref{eq_cost_function} in the main text as circles for different trajectory lengths estimated from $N=100$ simulated trajectories with the input as in (a), but with (b) $\sigma_{\rm{loc}}=0$, (c) $\sigma_{\rm{loc}}=0.9\sqrt{B}\Delta$ and (d) $\sigma_{\rm{loc}}=1.8\sqrt{B}\Delta$.
		The respective mean values of the two different estimates of $B$ for trajectory length $L/\Delta=4\times10^4$ are given in the upper right corner.
		The black dashed line represents the expected variance for uncorrelated data given in eq. \eqref{varB_uncorr} and the cyan line the expected variance given by eq. \eqref{eq_variance_estim_corr} using the analytical result for the correlation function given in eq. \eqref{eq_prw_vsqacf_analytic} with $n_{\rm{cut}}=n=L/\Delta-1$.
		As dictated by the law of large numbers, the variance is inversely proportional to the number of data points, i.e. inversely proportional to the trajectory length $L$.}
		\label{fig_appendix2}
	\end{figure*}
	
	In Fig. \ref{fig_appendix2}a we show extracted PRW parameter distributions from simulations with identical input parameters for different trajectory lengths $L$, where one can clearly see the variance decreasing for increasing trajectory length.
	In Figs. \ref{fig_appendix2}b-d this decrease is presented for different localization noise widths $\sigma_{\rm{loc}}$. The analytic prediction for uncorrelated velocities eq. \eqref{varB_uncorr} shown as black dashed line in Figs. \ref{fig_appendix2}b-d is clearly below the observed variance, since the velocities in the PRW are intrinsically correlated, but correctly describes the powerlaw dependence on $L$.
	For increasing localization noise, the relative variance $\sigma^2(B)/B^2$ decreases for $B_{\rm{est}}$ (crosses in Figs. \ref{fig_appendix2}b-d), because the mean increases by
	\begin{equation}
		\label{eq_B_noise}\langle B_{\rm{est}} \rangle = \frac{2C_{\rm{MSD}}^{\rm{theo}}(\Delta) + 4 \sigma_{\rm{loc}}}{2\Delta^2} \approx B + \frac{2\sigma_{\rm{loc}}}{\Delta^2}\,,
	\end{equation}
	which follows directly from eq. \eqref{eq_vacf_fit} of the main text.
	The approximation in the latter part of eq. \eqref{eq_B_noise} is accurate when the time step is smaller than the length of the ballistic regime in the MSD; in the case of the PRW, this means it is exact for $\Delta<\tau_m$.
	The localization noise increases the estimation of the mean squared velocity, as seen in eq. \eqref{eq_B_noise} and it leads to a decorrelation of consecutive velocities, as seen for instance in Figs. \ref{fig_3}e,f in the main text by the dip in the VACF at $t=\Delta$.
	Therefore, the estimator $B_{\rm{est}}$  eq. \eqref{eq_def_Bestimator} comes closer to the prediction for uncorrelated velocities for increasing localization noise, as seen by comparing the crosses in Fig. \ref{fig_appendix2}a to the crosses in Fig. \ref{fig_appendix2}b and c.
	
	The mean of $B_{\rm{est}}$ according to eq. \eqref{eq_def_Bestimator} is shown in the upper right corner of Figs. \ref{fig_appendix2}b-d, which increases with increasing localization noise as expected from eq. \eqref{eq_B_noise}.
	Here, we also show that $B$ determined by our fit of eq. \eqref{eq_cost_function}, denoted by $B_{\rm{fit}}$, gives accurate estimates close to the input value of $B$ even for high localization noise width $\sigma_{\rm{loc}}$.
	Moreover, the variance $B_{\rm{fit}}$, shown as circles in Figs. \ref{fig_appendix2}b-d, is close to the analytical estimate for $\sigma(B_{\rm{est}})$ according to eq. \eqref{eq_variance_estim_corr}, shown as the cyan line in Figs. \ref{fig_appendix2}b-d.
	However, this close match is not guarantied for arbitrary models, since the derivation cannot be done analytically for non-linear fit models, such as our fit by eq. \eqref{eq_cost_function}.
	Additionally, estimating the parameter variance of a trajectory ensemble of different trajectory lengths $L$ further complicates the matter.
	All these complications are avoided by estimating the parameter variance by simulating the underlying process based on the GLE eq. \eqref{eq_gle}, where all experimental settings, such as time step, different trajectory lengths and localization noise, can easily be included, as explained in the main text.

	\section{Test method with known origin parameter distribution}
\begin{figure*}
		\centering
		\includegraphics[width=\textwidth]{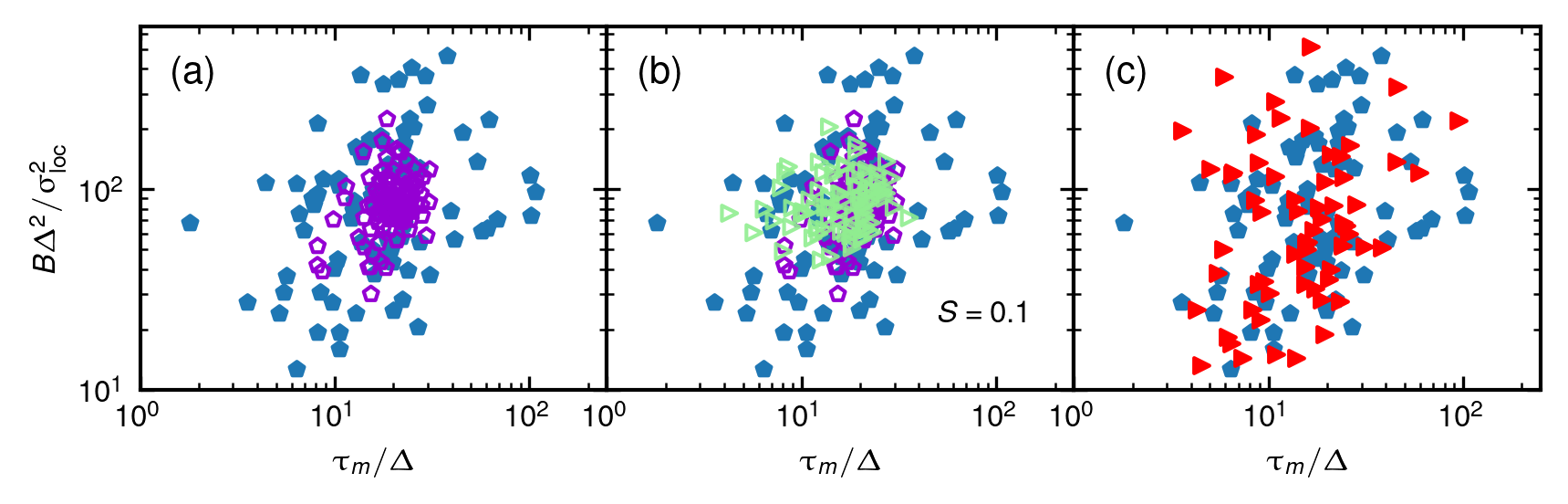}
		\caption{(a) $N=100$ randomly drawn PRW parameter vectors from a Gaussian distribution with covariance and mean given in eqs. \eqref{eq_cov_orig_test}, \eqref{eq_mean_orig_test} respectively, are shown as empty violet pentagons.
		Filled blue pentagons denote the parameters extracted from simulations using the empty violet pentagon input parameters for trajectories of length $L=100\,\Delta$.
		(b) A Gaussian distribution of empty green triangles with $S=0.1$ compares well to the original violet pentagon distribution that was used to create trajectories and extract the synthetic distribution of blue pentagons.
		(c) The red triangles represent the parameters extracted from the simulation using the green triangles in (b) as input parameters.}
		\label{fig_appendix1}
	\end{figure*}

	In order to show, that the statistical test explained in the Methods actually results in correct estimates for the cell-to-cell variance, we use a synthetic parameter distribution of known origin.
	For the PRW model eq. \eqref{eq_kern_prw} the parameter vector has three entries $\vec{z}=(\tau_m,\, B,\, \sigma_{\textrm{loc}})$, so we start by drawing $N$ parameter vectors from a Gaussian distribution with known covariance
	\begin{equation}
		\label{eq_cov_orig_test}
		\rm{Cov}^{\rm{inp}*} = 
	\begin{pmatrix}
		22.5 & 68.2 & 0.17 \\
		68.2 & 440.5 & -0.017 \\
		0.17 & -0.017 & -0.012
	\end{pmatrix}
	\end{equation}
	and mean
	\begin{equation}
		\label{eq_mean_orig_test}
		\langle \vec{z}\rangle=(10\unit{min}, 1.25\unit{\mu m^2/min^2}, 0.1\unit{\mu m})\,,
	\end{equation}
	where the mean is in the order of the parameters exhibited by the cancer cells as seen in Fig. \ref{fig_4}e.
	For clarity we omit the units of the covariance in eq. \eqref{eq_cov_orig_test}.
	A Gaussian distribution with covariance given by eq. \eqref{eq_cov_orig_test} and mean given by eq. \eqref{eq_mean_orig_test} is shown as violet empty symbols in Fig. \ref{fig_appendix1}a.
	We show the axes in Fig. \ref{fig_appendix1} in relative units because this way the results are better comparable across different systems with different units and the simulation results only vary if the ratios of parameters change.
	Now, we simulate a trajectory for each of the $N$ parameter vectors $\vec{z}$ of this known distribution of length $L=100\,\Delta$ with $\Delta=0.5\unit{min}$, which is a typical length for single-cell tracking data.
	
	In order to obtain the correct continuous time behavior, we simulate trajectories with a much smaller time step than the discretization of the trajectories with $h=0.05\Delta$ and then obtain trajectories by using only every $\Delta/h=$ 20th step of the simulation and adding localization noise drawn from the Gaussian distribution of width $\sigma_{\rm{loc}}$. For simulations of the algae cells we use $h=0.005\Delta$.
	The extracted parameters from these simulated trajectories are denoted by the blue symbols shown in Fig. \ref{fig_appendix1}a.
	Since we know the exact distribution that led to this extracted parameter distribution, we can now test whether our method reproduces the spread of the original distribution.
	The ratio $S$ defined in eq. \eqref{eq_result} of the input covariance $\rm{Cov}^{\rm{inp}*}$ to the covariance of the distribution extracted from the synthetic data, here denoted as $\rm{Cov}^{\rm{exp}}$, is $S = 0.11$.
	Indeed we find that the estimated covariance ratio using the statistical test introduced in the Methods is $S^* = 0.17\pm 0.08$.
	An exemplary distribution with $S=0.1$ is shown in Fig. \ref{fig_appendix1}b as empty green triangles and agrees well with the true original distribution shown as empty violet pentagons. Hence, the distribution extracted from simulations using the empty green triangles in Fig. \ref{fig_appendix1}b shown in Fig. \ref{fig_appendix1}c (red triangles) agrees with the distribution extracted from the synthetic data (blue symbols).
	
	We note that if the type of the underlying distribution is known, one can easily adapt the method to use an input distribution different from a Gaussian distribution.
	Nevertheless, in the most common case where the distribution type is not known, the Gaussian approach is most likely to describe the distribution.
	
	\begin{figure}
		\centering
		\hspace{-11mm}
		\includegraphics[width=0.53\textwidth]{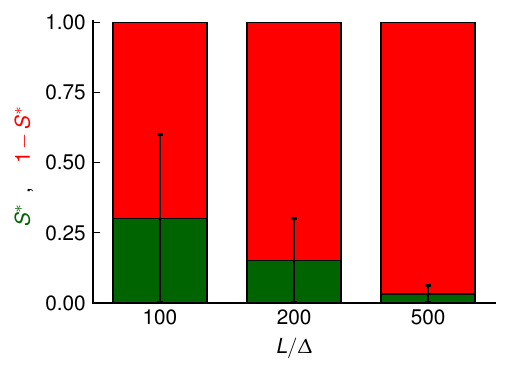}
		
		\caption{(a) Estimated variance ratio $S^*$ shown in green for synthetic distributions with all identical input parameters $B\Delta^2 /\sigma_{\rm{loc}}^2=80$, $\tau_m/\Delta=25$ ($S=0$) and different trajectory lengths $L$ for $N=100$ synthetic trajectories.}
		\label{fig_barplot_medsim_SI}
	\end{figure}
	
	Additionally, we create synthetic distributions of $N=100$ PRW trajectories originating all from the same parameter vector $B\Delta^2 /\sigma_{\rm{loc}}^2=80$, $\tau_m/\Delta=25$ close to the values of the breast cancer cell data.
	All parameters being the same means $\rm{Cov}^{\rm{inp}*}=0$ and hence the expected result is $S^*=0$. We create synthetic distributions by simulations of length $L=500\,\Delta$, $L=200\,\Delta$ and $L=100\,\Delta$ respectively and show the results for the estimated variance ratio $S^*$ in Fig. \ref{fig_barplot_medsim_SI}.
	All of the results agree with the expected value of $S^*=0$, but the uncertainty is higher for shorter trajectories, since the spread and the fluctuations of the fitted correlation functions are higher for shorter trajectories.

\makeatletter\@input{aux_file.tex}\makeatother